*Rare-Event Sampling: Occupation-Based Performance Measures for Parallel Tempering and Infinite Swapping Monte Carlo Methods*

J. D. Doll,[1] Nuria Plattner,[1,2] David L. Freeman,[3] Yufei Liu[4] and Paul Dupuis[4]

[1]Department of Chemistry
Brown University
Providence, RI 02912

[2]Department of Chemistry
University of Basel
Klingelberegstrasse 80
CH-4056, Basel Switzerland

[3]Department of Chemistry
University of Rhode Island
Kingston, RI 02881

[4]Division of Applied Mathematics
Brown University
Providence, RI 02912

*Abstract*

In the present paper we identify a rigorous property of a number of tempering-based Monte Carlo sampling methods, including parallel tempering as well as partial and infinite swapping. Based on this property we develop a variety of performance measures for such rare-event sampling methods that are broadly applicable, informative, and straightforward to implement. We illustrate the use of these performance measures with a series of applications involving the equilibrium properties of simple Lennard-Jones clusters, applications for which the performance levels of partial and infinite swapping approaches are found to be higher than those of conventional parallel tempering.

***I. Introduction:*** Monte Carlo methods[1-4] constitute an important and versatile set of tools for the study of many-body systems. By providing a refinable means for extracting macroscopic properties from specified microscopic force laws, such methods permit the atomistic study of systems of realistic physical complexity without the need for the introduction of uncontrollable approximations.

Although robust and general purpose, important practical issues can arise in the application of Monte Carlo techniques.[2-4] One such matter is the general problem of rare-event sampling. In typical equilibrium applications, where the relevant numerical task involves estimating averages of interest over known probability distributions, the adequacy of the sampling methods involved is an obvious and critical issue. If the probability distribution has a single, simply-connected region of importance, ordinary random walk sampling procedures[1,3-5] are generally adequate. If, however, the distribution in question contains multiple, isolated regions of importance, transitions between them can become infrequent ("rare") when using conventional Metropolis-style methods rendering the associated property estimates unreliable. Unfortunately, from a practical point of view, applications in which such rare-event difficulties arise are themselves not rare. They arise frequently, for example, in studies of activated processes,[6] applications of substantial importance in chemical, biological and materials investigations.

A number of approaches have been devised in an effort to overcome rare-event sampling concerns. One of the more widely used is the parallel tempering[7-8] (PT) or replica exchange technique.[9] As summarized elsewhere,[6] this method utilizes an expanded computational ensemble composed of systems corresponding to a range of control parameters such as the temperature. The core idea is to make use of information produced by one portion of the ensemble (e.g. higher temperatures) to improve the sampling in another (e.g. lower temperatures). The information transfer needed to improve the sampling is generally accomplished by augmenting conventional Metropolis-style particle displacements with suitably designed "swaps" of coordinates between ensemble temperatures. Strategies for the selection of the ensemble temperatures[10-14] as well as for the frequency[15,16] and nature[17-20] of the swap attempts have been discussed.

Recently, a large deviation analysis of the performance of parallel tempering has led to the development of a new class of rare-event methods, the infinite swapping (INS) approach.[21,22] In its most complete form the INS method can be viewed as the extreme limit of parallel tempering in which swaps involving all possible temperatures are attempted at an infinitely rapid rate, a limit the large deviation analysis demonstrates to be optimal. The infinitely rapid swaps induce a Born-Oppenheimer like environment, one in which the relevant distribution becomes a thermally symmetrized analog of that used in conventional parallel tempering. Although the computational requirements for the full INS method grow rapidly with the number of temperatures involved, practical methods that capture a substantial level of the performance potential of the full approach while offering a significant cost/performance increase relative to parallel tempering are available. Details of these partial infinite swapping (PINS) approaches and their implementation are discussed elsewhere.[21,22]

Performance measures are important tools in the development and application of rare-event sampling methods. Such measures are necessary, for example, for making decisions concerning questions ranging from whether or not a particular method is "working" to those related to the relative performance of alternative approaches or implementations. Ideally, the measures in question should be informative, straightforward to implement and of sufficient generality that their utility transcends specific systems or properties. In practice, since approaches may differ with respect to these desirable features, having a variety of performance measures available is useful.

One general strategy for approaching the discussion of performance measures is to identify criteria that are plausibly related to the functioning of the computational method involved and to utilize those criteria to "optimize" the associated simulation. In the case of the parallel tempering approach, for example, tempering swaps are at the core of the method and are thus a reasonable focus for attention. Kofke has posited that achieving a uniform acceptance probability for such swap attempts across the computational ensemble is a desirable goal, one that provides a means for selecting ensemble temperatures.[10,23] In their work Predescu and co-workers[11,12] have expanded this line of argument by linking the acceptance probabilities involved to system heat capacity information. A somewhat different approach, explored by Katzengrabber, et al.,[14] is based on the premise that in a parallel tempering application it is the *rate* at which configurations transit the computational ensemble that is of primary interest. Applications have shown that the optimal tempering ensembles produced by the uniform acceptance and maximum rate approaches are in general not identical.

Rather than beginning with the identification of an optimization criterion, another approach in the discussion of performance measures is to identify a property (or properties) inherent in the simulation method itself and to utilize the presence/absence/rate of achievement of this property as a performance measure. Neirotti, et al.[24] adopt such an "inherent property" approach in their study of rare-event sampling. Building upon earlier work by Thirumalai and co-workers,[25] they utilize the decay of energy-related metrics to known limiting values to monitor sampling.

In the present paper we explore the development and application of occupation-based approaches for tempering-based Monte Carlo methods. The outline of the paper is as follows. We begin in Section II with an empirical observation that points to a somewhat surprising property of parallel tempering methods. After demonstrating the rigorous and general nature of this result, we use the property involved to develop a number of convenient performance diagnostics for both parallel tempering and partial swapping methods. In Section III we illustrate the use of the resulting diagnostics for a number of numerical examples of varying complexity that involve the equilibrium properties of models of simple rare-gas clusters. Section IV contains a summary and discussion of our results.

*II. Background and An Observation:* We begin by considering a number of tempering investigations of a particular system, a Lennard-Jones model of a thirteen atom rare-gas cluster. In addition to providing a concrete framework for the present discussion, these simple cluster simulations contain hints of features that prove useful in the broader consideration of rare-event sampling. We find it convenient to express the results with reference to a particular system (argon) thereby giving a physical context to the observed features noting that computed properties for the Lennard-Jones model are universal and can be expressed using reduced variables.

Occupation traces of the type introduced by Katzengrabber, et al.[14] are a convenient device for discussing tempering based sampling methods. Such traces are a record of the temperature associations for a particular coordinate set as it moves within the tempering ensemble during the simulation. For parallel tempering applications there are clear temperature-coordinate associations at each step in the simulation. For infinite and partial swapping approaches, on the other hand, such explicit associations are obscured by the symmetrization involved. Although this makes the identification and construction of occupation traces for INS and PINS applications a somewhat more subtle issue, a generalization of the basic parallel tempering result proves both possible and practical. The key is to base such a generalization on steps in the sampling process for which the temperature-coordinate associations are unambiguous.

Sampling in the present work is based on the "dual-chain" approach, a method that is described in detail in Refs. (21-22). This technique involves partitioning the ensemble temperatures into contiguous blocks in two distinct ways ("dual chains"). Although symmetrization in this approach is "partial" (i.e. occurs only within the various separate temperature blocks), by suitably combining moves within the individual chains with others that exchange information between them it is possible to produce a sampling of the fully symmetrized, infinite swapping distribution. Most importantly, this sampling is accomplished without the factorial-scale growth in computational effort that would be incurred in a brute-force INS approach. The "hand-off" or transfer of information between the two sampling chains is a critical step in the dual-chain process. During such hand-offs explicit temperature-coordinate associations are established, associations that provide a natural basis for the construction of INS and PINS occupation traces and that are utilized for such tasks in the present work.

Figure (1) shows a number of occupation traces obtained from short, three-temperature simulations of the $Ar_{13}$ cluster. The traces displayed chronicle the progress of configurations that are initially at the lowest of the three ensemble temperatures for a period of 1000 moves. Analogous traces for configurations initially associated with either of the other two ensemble temperatures could be similarly constructed. The low and high-temperatures used in all simulations in Fig. (1) are $T_1 = 30$ K, $T_3 = 40$ K, respectively, values chosen to bracket the temperature of the $Ar_{13}$ cluster's heat capacity maximum (34 K).[26] The intermediate temperature, $T_2$, varies being 35 K in Figs. (1a) and (1b) and 39 K in Figs. (1c) and (1d). Figures (1a) and (1b) correspond to PT simulations while Figs. (1c) and (1d) are analogous, three-temperature PINS results. Unless otherwise stated all numerical simulations in the present work utilize the methods described in Appendix A and in Refs. (21-22). The dual chains in the PINS simulations have one chain with temperature blocks that consist of $T_1$ and $T_{2-3}$ and another chain with blocks that consist of $T_{1-2}$ and $T_3$. We see in Fig. (1) that the movement of the system in question throughout the computational ensemble is sensitive to both ensemble choice ($T_2$ value) and to sampling method (PT or PINS).

Trebst, et al.[14,27] have suggested that the number of sampling moves required to traverse the computational ensemble provides a convenient measure of sampling performance. This idea is illustrated for the present application in Fig. (2) where the average number of moves required for the system to make a round trip across the various computational ensembles, $<n_{rt}>$, is plotted as a function of the choice of the intermediate ensemble temperature, $T_2$, for the various methods. These results are obtained from occupation traces that contain a total of $2^{17}$ moves. As anticipated, there is a minimum in $<n_{rt}>$ values as a function of the choice of $T_2$ for the various sampling methods. The "optimal" $T_2$ choice for this system (i.e. the value that produces the most rapid traversal of the computational ensemble) is approximately 35 K for both PT and PINS methods. For a specified value of $T_2$, on the other hand, we see in Fig. (2) that the movement across the computational ensemble is appreciably more rapid with the INS and PINS methods than with parallel tempering. For example, for $T_2 = 35$ K the PT, PINS and INS $<n_{rt}>$ values are approximately 298.6, 61.2 and 29.9, respectively. As an aside, the "flatter" nature of the INS and PINS plots in Fig. (2) suggests that the performance of such methods is more robust than that of parallel tempering with respect to sub-optimal choices of the computational ensemble.

We now turn to a somewhat unexpected aspect of the tempering results in Fig. (1). Beyond its intrinsic interest, this feature provides the basis for a number of simple, property-independent tools that prove useful in the characterization and analysis of rare-event sampling issues. Table I examines the fraction of moves various three-temperature $Ar_{13}$ occupation traces of the type in Fig. (1) spend in the different temperature streams. The occupation traces used to generate these results are longer ($2^{23}$ total moves), but otherwise of the type shown in Fig. (1). The numerical results of Table I suggest that the fractional occupancies of the various temperature streams are in fact *uniform* for *all* the tempering methods involved *regardless of the choice of $T_2$*. Such uniformity turns out to be a general and rigorous property of INS, PINS and parallel tempering approaches and holds for an arbitrary number of temperatures. That is, if there are $N_t$ temperatures in the computational ensemble, the fraction of the total number of moves such occupation traces spend in any particular ensemble temperature asymptotically approaches a value of $1/N_t$. This "equal occupancy" result is one of the principal findings of the current work.

Although the property appears not to have been noticed previously, it turns out that the equal occupancy result suggested by Table I is a general feature of the parallel tempering method. We provide a heuristic argument for this claim below. A more detailed argument as well as extensions to cover both infinite and partial swapping approaches are given in Appendix B.

We begin by considering an M-step parallel tempering simulation that involves $N_t$ temperatures, $\{T_n\}$, n=1,$N_t$. As they move about in the computational ensemble, the initial configurations in such a simulation generate $N_t$ distinct occupation traces, one for each of the possible initial temperatures. If we designate the temperatures for one such trace, labeled α, at step m in the simulation as $T_\alpha(m)$, then the fraction of moves trace-α spends in temperature stream $T_n$, $f_n^\alpha$, is given by

$$f_n^\alpha = \frac{1}{M} \sum_{m=1}^{M} \mathbf{1}_{T_\alpha(m), T_n},$$

(2.1)

where

$$\mathbf{1}_{T_\alpha(m), T_n} = \begin{cases} 1, & T_\alpha(m) = T_n \\ 0, & otherwise \end{cases}.$$

(2.2)

The underlying occupation traces are stationary, ergodic random processes. Consequently, the asymptotic statistical properties of the $N_t$ possible occupation traces arising from the various starting locations must be equivalent (i.e. there can be no dependence of the $f_n^\alpha$ values on the index α). As a result the values of $f_n^\alpha$ for a given temeprature approach a common value, $f_n$, for all $N_t$ possible occupation traces. Since we know that the $f_n^\alpha$ values are all asymptotically equal for all values of α, and since we know from Eq. (2.1) that their sum over α for a given temperature totals unity, we conclude that $f_n = 1/N_t$ for any temperature $T_n$. Other than assuming it to be non-zero, the equal occupancy result does not depend on a particular choice for the parallel tempering swap attempt frequency. As noted in the Introduction, the infinite swapping approach can be viewed as a limiting form of parallel tempering in which swaps involving all possible temperatures are attempted at an infinitely rapid rate. One thus suspects (properly) that the parallel tempering equal occupancy result is also valid for INS and PINS approaches. A demonstration of this general result is presented in Appendix B.

The uniform occupation property of the parallel tempering, PINS and INS methods provides the basis of a number of useful sampling performance measures. One such measure is the Shannon entropy associated with the occupation fractions for the simulation. If the occupation fraction for temperature stream $T_n$ after $n_{move}$ simulation moves is designated by $f_n(n_{move})$, then the occupation entropy for the simulation, $S_f$, is defined as

$$S_f = -\sum_{n=1}^{N_t} f_n(n_{move}) \ln[f_n(n_{move})].$$

(2.3)

When $n_{move} = 1$, $f_n(n_{move}) = \delta_{n,s}$, where s is the index of the initial temperature stream of the trace. So defined the $S_f$ value begins at zero and increases as the simulation proceeds. Because the asymptotic limit of $S_f$ is known (i.e. $S_{max} = \ln(N_t)$), both the *convergence* and the *rate of convergence* to this limit can serve as practical performance diagnostics. At the crudest level the failure of $S_f$ to achieve its known limiting value signals an obvious breakdown in the sampling. More generally, the rate of approach of $S_f$ to this limit provides a convenient, quantitative, property-independent means for comparing the performance of different computational ensembles and/or sampling methods.

Figures (3) and (4) contain plots of the entropies computed from the PT and PINS occupation traces for $Ar_{13}$ from Table I. We see in Fig. (3) that the entropies for all simulations considered appear to achieve their proper limiting values, which in this case is ln(3). We also see in Fig. (4) that the rate at which this limit is achieved for a specified intermediate temperature is more rapid with the PINS method than with parallel tempering. In fact, judged by this entropic measure the results of Fig. (4) indicate that the PINS performance for a "bad" choice of intermediate temperatures ($T_2 = 39$ K) is actually superior to that of parallel tempering for the optimal choice ($T_2 = 35$ K).

A related performance measure is the fluctuation autocorrelation function of the occupation trace index. If we define the index of the temperature at step m in an occupation trace as N(m), where $1 \leq N \leq N_t$, and the fluctuation of this value about its overall average as $\delta N(m)$, then the correlation function, C(s), for the stationary random process $\delta N(m)$ is defined as

$$C(s) = \frac{\langle \delta N(m) \delta N(m+s) \rangle}{\langle \delta N(m) \delta N(m) \rangle},$$

(2.4)

where the brackets in Eq. (2.4) denote an average over the occupation trace in question. Plots of C(s) for the various occupation traces of Table I are shown in Fig. (5). We see that for this rather simple application the performance rankings suggested by Fig. (5) are consistent with those in Figs. (3-4) as well as with those based on the round-trip transit measures shown in Fig. (2).

Although modest in the examples considered thus far, the utility of the occupation-based performance measures set forth in Eqs. (2.3) and (2.4) can become more significant when dealing with more challenging applications. An example is illustrated in Figs. (6) and (7). Figure (6) shows occupation traces for a 66-temperature $Ar_{38}$ PINS and PT simulations, both of which begin in the highest temperature. The details of these and related simulations will be described in Section III. For the moment the relevant feature to note in Fig. (6) is that over the course of the simulation the PINS occupation trace moves throughout the entire computational ensemble making a total of nine round trips across the 10-30 K interval, a range that includes both solid-like and liquid-like cluster behavior, while PT trace's movement is appreciably more limited. The failure of the PT occupation trace to complete even a single transit of the computational ensemble during the simulation interval shown is sufficient to raise questions concerning the adequacy of its associated sampling. In contrast, the ease and frequency with which the PINS occupation trace in Fig. (6) transits the ensemble could, if taken by itself, be viewed as evidence that all was in order and that the properties derived from the underlying simulation are reliable. The occupation analysis in Fig. (7), however, indicates that such a conclusion is premature. In particular, we see in Fig. (7) a significant *underrepresentation* of visits to the low-temperature portions of the ensemble. The occupation entropy for the distribution shown in Fig. (7) is $S_f = 3.98$, well short of the $\ln(66) = 4.19$ uniform limit. As discussed more fully in Section III, the underrepresentation of low-temperature visits seen in Fig. (7) is ultimately eliminated as more sampling points are included. The point we wish to emphasize, however, is that this underrepresentation is an indication of an important deficiency in the original, short simulation, a deficiency detected by the equal occupancy analysis, but not by transit-based measures.

In what follows it is useful to recast the difference between the actual and maximum occupation entropies in slightly different terms. Specifically, if we view the calculated $S_f$ value as arising from a uniform limit of a fictitious ensemble, the "active number" of temperatures in that ensemble, $N_a$, would be the exponential of the associated occupation entropy. In the case of Fig. (7) $N_a$ would thus be exp(3.98) or 53.5, appreciably smaller than the actual number of 66. We can also define a related quantity, the "active fraction" for the simulation, $F_a$, as the ratio of $N_a$ to the actual number of temperatures, $N_t$. In the example associated with Figs. (6) and (7), $F_a$ is 53.5/66, or approximately 0.810. The concepts of active ensemble size and fraction will prove convenient in later discussions.

*III. Illustrative Applications:* To explore further the utility of the performance measures developed in Section II, we consider their application to systems in which the number of temperatures in the computational ensemble and the physical complexity of the systems involved are increased.

Our first example is designed to illustrate how one can utilize the performance measures discussed in Section II to aid in the selection of both the tempering ensemble and the sampling method. Figures (8-11) contain various occupation entropy and autocorrelation function results for the $Ar_{13}$ cluster. These are obtained with three different computational ensembles using both the PINS and PT methods outlined in Section II and Appendix A. The computational ensembles involved are each composed of a common number of temperatures, $T_1$ - $T_{24}$, with $T_1 < T_2, ...., < T_{24}$ and $T_1$ = 20 K, $T_{24}$ = 50 K. Although they span the same overall range, the ensembles are chosen to have different temperature distributions. Specifically, in ensemble-a temperatures $T_1$-$T_{12}$ and $T_{13}$-$T_{24}$ uniformly cover the intervals 20-28 K and 42-50 K, respectively while in ensemble-b the two, uniform twelve-temperature regions range from 20-32 K and 38-50 K. The third ensemble, ensemble-c, consists of 24 temperatures uniformly distributed over the entire 20-50 K interval. The PINS results are obtained using dual-chain sampling methods of the type discussed previously. In these simulations one chain is composed of four symmetrized blocks of six temperatures, ($T_{1-6}$, $T_{7-12}$, $T_{13-18}$, $T_{19-24}$), while the other chain consists of five symmetrized blocks, ($T_{1-3}$, $T_{4-9}$, $T_{10-15}$, $T_{16-21}$, $T_{22-24}$). The PINS simulations utilize $2^{16}$ total moves while the PT results utilize $2^{18}$. The other computational details for the PINS and PT simulations are discussed in Section II and Appendix A.

We see in Figs. (8-9) that for a given ensemble the rate of convergence of the PINS occupation entropies to the uniform, asymptotic limit of ln(24) is appreciably greater than that achieved with parallel tempering. We also see that the asymptotic convergence rate of the PINS results is highest for ensemble-c of the three considered. Interestingly, the *initial* (small $n_{move}$) PINS convergence rates seen in Figs (8-9) are similar for all three ensembles. As seen most clearly in Fig. (8), the PINS occupation entropy increase for ensemble-a is rapid up to a value of approximately $S_f = 2.5$ after which a distinctly slower rate of increase is observed. This suggests ($\exp(2.5) \approx 12$) that equilibration is rapid within the 12-temperature blocks of ensemble-a, but is ultimately hindered by the large temperature "gap" between 28-42 K in this ensemble. As with the analogous PINS results, the rate of approach of the occupation entropy to its uniform limit for the PT results in Fig. (8) is slowest for ensemble-a. While not identical, PT results for ensemble-b and ensemble-c are numerically quite similar for this system.

Figure (10) contains plots of the fluctuation autocorrelation functions, C(s), obtained from the various 24-temperature $Ar_{13}$ simulations shown in Figs. (8-9). The plots in Fig. (10) reinforce the basic convergence story conveyed by the entropic measures and serve to illustrate the dramatic difference in the decay rates seen in the PINS and parallel tempering simulations. This difference is also conveyed by Fig. (11) where we plot a small portion of the PINS and PT occupation traces obtained for the best of the three 24-temperature computational ensembles, ensemble-c. Although both PINS and PT traces show frequent passage through the temperature region of the heat capacity maximum (34 K), the qualitatively different character of the two traces provides a visual sense of the far greater rate of information transfer within the computational ensemble achieved by the PINS approach relative to the PT method for this application. This dramatic difference in the movement of the PINS and PT traces through the computational ensemble lies behind the vastly different rates of decay of the corresponding PINS and PT fluctuation autocorrelation functions seen in Fig. (10).

We now turn to our second, more challenging example, the 38-atom Lennard-Jones cluster. The LJ-38 system exhibits an interesting and diverse phenomenology, and, as a consequence, has received a significant amount of attention.[28] As discussed in the pioneering work of Doye, Miller and Wales,[29] this cluster has a double-funnel potential energy landscape for which the global minimum and the lowest lying local minimum are similar in energy, dissimilar in structure, and separated by a relatively large energy barrier. In particular, the global minimum and lowest local minimum correspond to fcc truncated octahedral and incomplete icosahedral structures, respectively.

We begin by specifying the computational ensemble. Previous work[18] has shown that the heat capacity of the LJ-38 system has a primary peak in the reduced temperature range of $k_BT/\varepsilon = 0.165$, a value that for argon corresponds to a physical temperature of approximately 19.8 K. As in the $Ar_{13}$ simulations discussed earlier in this section, we choose the overall temperature range for our $Ar_{38}$ simulations to bracket the temperature of the major heat capacity maximum. Here we choose to include temperatures that cover the range $10 \text{ K} \leq T \leq 30 \text{ K}$, a region previous structural studies have shown marks the transition from solid-like to liquid-like behavior in the $Ar_{38}$ cluster. The choice of the number of temperatures to include in the ensemble is somewhat arbitrary. In general, we want the number to be large enough to facilitate information flow within the ensemble, but small enough to make the simulation computationally manageable. Based on considerations described in Appendix C, we utilize a 66-temperature ensemble. The required PINS sampling is performed using the dual-chain sampling methods described in Refs. (21-22). Assuming that the temperatures involved are ordered, $T_1 < T_2 < ....$, $< T_{66}$, one chain consists of eleven symmetrized blocks of six temperatures each ($T_{1-6}$, $T_{7-12}$,....,$T_{61-66}$), while the other consists of 10 symmetrized blocks of six temperatures each that form the chain's interior and two symmetrized blocks of three temperatures that form the low and high-temperature caps for the chain, ($T_{1-3}$, $T_{4-9}$,....,$T_{58-63}$,$T_{64-66}$). The specific values of $T_1$-$T_{66}$ are listed in Appendix C.

Figures (12-13) contain a variety of results for the $Ar_{38}$ cluster obtained from two PINS simulations based on the 66-temperature ensemble described in Appendix C. Each of the two simulations contains a total of 2.0 x $10^6$ moves of which the last 1.6 x $10^6$ are utilized for data collection. To check that the computed results are independent of the starting conditions, one simulation is initiated using configurations corresponding to the cluster global minimum geometry while the other is initiated using configurations corresponding to the lowest-lying icosahedral local minimum. Figures (12) and (13) show the occupation entropies (c.f. Eq. (2.3)) obtained from the two $Ar_{38}$ simulations. Both PINS simulations converge to their uniform limits in a similar fashion. For comparison, analogous parallel tempering results are also presented in Fig. (12). The PT simulation uses the same 66-temperature ensemble and underlying smart Monte Carlo displacement strategy used in the PINS studies. A significant difference in the levels of performance for the PINS and PT approaches is apparent.

In Fig, (14) we show plots of the temperature dependence of the average structural order parameter, $<Q_4(T)>$, obtained from the two PINS simulations associated with Figs. (12) and (13). $Q_4$ is a convenient choice for distinguishing structures associated with the two major potential energy funnels of the LJ-38 system, its value ranging from near zero for icosahedral and melt structures to approximately 0.19 for fcc-like forms. Used by a variety of investigators,[29,30,18] these order parameters are defined more generally as

$$Q_\ell = \left( \frac{4\pi}{2\ell+1} \sum_{m=-\ell}^{\ell} |\overline{Q}_{\ell m}|^2 \right)^{1/2},$$

(3.1)

where

$$\overline{Q}_{\ell m} = \frac{1}{N_b} \sum_{r_{ij}<r_0} Y_{\ell m}(\vartheta_{ij}, \phi_{ij}),$$

(3.2).

and where the summation in Eq. (3.2) is over all pairs of atoms ("bonds") for which the separation distance, $r_{ij}$, is less than a preset threshold, $r_0$. In line with previous studies,[18,29] the $r_0$ value used in the current work is taken to be 1.391 times the Lennard-Jones σ-parameter. Again we see equivalence in Fig. (14) for the results produced by the fcc and icosahedrally-initiated simulations.

We close by returning to the issue raised at the end of Section II, the breakdown in the equal occupancy behavior seen in Fig. (7). Figure (15) shows the instantaneous $Q_4$ values for the configurations visited during a somewhat longer segment of the occupation trace presented in Fig. (6). During the portion of the trace shown in Fig. (6) ($n_{move} < 10^5$), we see $Q_4$ values representative of those for the icosahedral and melt forms of the cluster, but not for those of the fcc global minimum energy structure. Because the global minimum exists preferentially at low-temperatures, its absence from the shorter occupation trace shown in Fig. (6) is the reason for the breakdown of the equal occupancy behavior seen in Fig. (7). It is interesting to note that in this example the equal occupancy measure is thus detecting the *absence* of visits to the global minimum. For completeness, we note that the final occupation entropies for the entire, 1.6 x $10^6$ move global minimum and icosahedrally-initiated PINS studies associated with Figs. (12)-(14) are 4.1875 and 4.1895, respectively. In the language of Section II, such occupation entropy values correspond to active ensemble fractions of 0.9978 and 0.9998, values that indicate equal occupancy is indeed achieved in the two simulations.

It is useful to examine the issue of equilibration raised by Fig. (15) from a somewhat different perspective. At any given instant in the PINS simulations one can evaluate the $Q_4$ values for the configurations in the various data streams of the computational ensemble. Such a set of values provides a rough sense of population in the various structural basins. Shown in Fig. (16) is the number of systems (out of the total of 66) for which the associated $Q_4$ values lie below a specified threshold (here taken to be 0.09) for the $Ar_{38}$ PINS simulations initiated in the global minimum and icosahedral basins. This number, $K(n_{move})$, is basically a count of the number of icosahedral and melt forms. It builds from an initial value of zero as the global minimum initiated simulation proceeds while the analogous, icosahedrally initiated result declines from an initial value of 66.

After roughly 400,000 moves the values of $K(n_{move})$ for the two simulations stabilize at an average value of approximately 45.

Shown in Fig. (17) are block averages of $Q_4$ values for the two PINS simulations at two specific temperatures, $T_1 = 10$ K and $T_{20} = 14.9350$ K. $T_1$ is the lowest ensemble temperature while (c.f. Fig. (15)) $T_{20}$ is the temperature for which $<Q_4>$ is roughly 0.1, the approximate mean of $Q_4$'s limiting values. As with the results in Fig. (17), we see that after an initial equilibration period the block averages for the two PINS simulations approach common limiting values. Interestingly, we see in Fig. (17) that the warm up period required to achieve stable estimates of the $<Q_4>$ values matches that seen in Fig. (16). In other words, the warm up period necessary to produce stable estimates of thermally averaged properties is dictated by the period required to establish the relative populations of the major energy basins. Parenthetically, we note that the equilibration period seen in Fig. (17) is the basis for our choice of 400,000 move warm up periods in the PINS $Ar_{38}$ simulations of this section. The values of $<Q_4>$ for each of the temperatures shown in Fig. (14) are constructed from the block averages of the type shown in Fig. (17) using the last $1.6 \times 10^6$ move portions of the two PINS simulations.

Results of the type shown in Fig. (15) represent simulation histories obtained by following a specified occupation trace throughout the simulation. Figure (18) represents another type of simulation history, one that follows a property associated with configurations that correspond to a *particular temperature*. Shown in Fig. (18) are the $Q_4$ values of $Ar_{38}$ configurations for a fixed temperature of 22.727 K ($k_B T/\varepsilon = 0.190$). The results in Fig. (18) are taken from a 10,000 move segment of the larger, icosahedral minimum initiated PINS simulation after the 400,000 move warm up period has been completed. Analogous results have been obtained for PINS simulations initiated using the global minimum configuration. These are statistically indistinguishable from those of Fig. (18) and are not shown. The particular temperature of 22.727 K is chosen to permit a comparison of the results of Fig. (18) with those of conventional Monte Carlo studies (c.f. Fig. (11) of Ref. (29)). Rapid switches of configurations with differing $Q_4$ values at a fixed temperature in the PINS simulation are evident.

***IV. Discussion and Summary:*** The ubiquitous and difficult nature of rare-event sampling issues makes the development of methods for their detection and treatment an important topic. In the present paper we have introduced a variety of occupation-based performance measures for tempering applications and have explored their utility with a number of numerical examples.

The equal occupancy result of Section II is the core of the present developments. Briefly summarized, equal occupancy for a tempering simulation implies that the asymptotic fraction of the total number of moves a particular configuration spends in a given data stream in a simulation that contains $N_t$ temperatures is a constant, $1/N_t$, independent of the choice of ensemble temperatures. Starting with the empirical observation of such behavior in parallel tempering simulations, we have demonstrated that the result is a general property of both parallel tempering as well as of partial and infinite swapping approaches.

Using the equal occupancy property we have constructed a number of associated performance measures and have explored their utility for applications involving models of simple, single-component atomic clusters. We have found the resulting methods both simple to implement and of appreciable utility. Based on these findings we have concluded that the occupation based performance measures presented represent useful tools in ongoing efforts involving the development and application of rare-event sampling methods.

*Acknowledgments:* The authors gratefully acknowledge grant support of this research through the DOE Multiscale and Optimization for Complex Systems program No. DE-SC0002413. NP wishes to thank the Swiss National Science Foundation for postdoctoral support and JDD wishes to acknowledge support through DOE departmental program No. DE-00015561. PD wishes to acknowledge support from the Army Research Office (W911NF-12-1-0222) and PD gratefully acknowledges support from the National Science Foundation (DMS-1008331). The authors also wish to thank Dr. James Gubernatis, Los Alamos National Laboratory for helpful discussions concerning the present work.

*Appendix A:* In this Appendix we describe the generic numerical methods used in the present studies including the underlying force laws and numerical methods.

The atomic-level force law in the present work is assumed to consist of pair-wise Lennard-Jones interactions between the atoms plus a center of mass confining potential to prevent cluster evaporation. Specifically, the total potential energy, U(**r**), for the N-particle cluster is given by

$$U(\mathbf{r}) = U_{LJ}(\mathbf{r}) + U_c(\mathbf{r}),$$

(A.1)

where

$$U_{LJ}(\mathbf{r}) = 4\varepsilon \sum_{i<j}^{N} \left( \left(\frac{\sigma}{r_{ij}}\right)^{12} - \left(\frac{\sigma}{r_{ij}}\right)^{6} \right),$$

(A.2)

and where

$$U_c(\mathbf{r}) = \varepsilon \sum_{i=1}^{N} \left( \frac{|\mathbf{r}_i - \mathbf{r}_{cm}|}{R_c} \right)^{20}.$$

(A.3)

The Lennard-Jones length and energy parameters, $\sigma$ and $\varepsilon$, are taken to be $\varepsilon = 119.8$ K and $\sigma = 3.405$ Å for argon. In the confining potential definition, $\mathbf{r}_{cm}$ is the center of mass of the cluster and $R_c$ is an empirical constant. The $R_c$ values used in the present work are $2.5\sigma$ and $2.65\sigma$ for the $Ar_{13}$ and $Ar_{38}$) investigations, respectively.

All partial swapping simulations in the present work are generated using the dual-chain sampling techniques described in detail in Appendix B of Ref. (21). Unless otherwise stated, the chains in the simulation are constructed in the following manner. One of the chains involved is composed entirely of blocks of six adjacent temperatures, $T_{1-6}$, $T_{7-12}$, etc. In the second chain, the first and last block are composed of three-temperatures while the remaining blocks have six temperatures each. The basic idea is to utilize symmetrized blocks that are large enough to promote appreciable mixing of temperature information, small enough to be computationally manageable, and overlap sufficiently to promote chain-to-chain information transfer.

Smart Monte Carlo (SMC) techniques[31] are utilized to perform the necessary, single-temperature sampling moves in this approach. For the present argon studies each such SMC move consists of a 100 step molecular dynamics segment (each step of $10^3$ au duration) in which the initial momenta are selected at random from equilibrium Boltzmann distributions of the appropriate temperatures. To provide a common basis for comparison, the parallel tempering simulations reported here also utilize the same SMC methods. Unless otherwise stated, all parallel tempering results are based on simulations in which at each step there is a 16% probability of an attempted swap of configurations between a randomly selected adjacent pair of temperatures.

The performance measures discussed in Section II are generated by suitably processing the occupation traces involved. Such occupation traces, defined in Section II, are a chronicle of the temperature index for the system in question at the $m^{th}$ step in the simulation, N(m). Once obtained, they are processed to produce the associated performance measures. Computing the fluctuation autocorrelation function for a specified occupation trace (c.f. Eq. (2.4)) is entirely analogous to the task of constructing time correlation functions from molecular dynamics simulation data and thus needs no further discussion. Computing the occupation entropy (c.f. Eq. (2.3)) requires knowledge of the fractional occupancies of the various tempering data streams. To compute the occupation entropy, $S_f(n_{move})$, associated with a trace that begins in a particular temperature stream, $n_0$, after a specified number of moves, $n_{move}$, for example, we need to know the relevant fractional occupancies of the various levels, $\{f_n(n_{move})\}$, n=1,$N_t$ produced by that trace. Instead of focusing on a particular starting point, $n_0$, it is generally more convenient to form the average of such individual occupation entropies over traces that originate in all possible starting locations. Unless otherwise noted, all occupation entropies in the present work are of this averaged type. It should also be noted that such averaged values can be computed from a single, long occupation trace by exploiting the stationary nature of the occupation traces involved.

*Appendix B:*  In this appendix we give two arguments to demonstrate the asymptotic equal occupancy of the temperature for PT, INS and PINS. The first argument formalizes the one used in Section II. Consider a stationary, ergodic Markov process (S(m), Y(m)) such that the first component takes values in the set $\mathcal{P}\{N_t\}$ of all permutations of $\{T_1,...,T_{N_t}\}$. PT falls into this framework when the pair of temperatures for which a swap is attempted is selected according to a randomized rule that does not depend on the past of the simulation and which allows all permutations to be realized. INS also falls into this framework where the S(m) component arises through an explicit temperature-coordinate association that can be made at each time step. If the selection of partitions in a PINS scheme is done on a randomized basis, then an explicit temperature-coordinate association must be done when switching between different partitions, and this again gives a representation for the scheme in the form described above.

With these definitions one can interpret S(m) (in the notation of Section II) as $(T_1(m),...,T_{N_t}(m))$. Since S(m) is a permutation, for each m = 1,2,... and each $T_n \in$ $\{T_1,...,T_{Nt}\}$ there is one and only one $\alpha$ such that $T_\alpha(m) = T_n$. With $f_n^\alpha(M)$ defined as in Section II, the ergodic theorem then implies the almost sure limit

$$f_n^\alpha(M) \to f_n$$

(B.1)

as $M \to \infty$, where $f_n$ is independent of $\alpha$, M, and Y(0). However, the fact that each n is identified with one and only one $\alpha$ implies that for each n and all M

$$\sum_{T_n \in \{T_1,...,T_{N_t}\}} f_n^\alpha(M) = 1,$$

(B.2)

from which $f_n = 1/N_t$ follows.

While this argument applies to stationary and ergodic implementations of PT and PINS, in practice they are often implemented in a nonstationary way. For example, in PT one may cycle repeatedly through the sequence of adjacent temperature pairs, while in the PINS scheme one could simply alternate between two different partitions (the "dual-chain" method used in Refs. (21) and (22)). While each such algorithmic step corresponds to a Markov transition kernel on (S,Y), the kernels differ (e.g., depending on the particular pair of temperatures in PT or the partition of temperatures used in PINS). Thus the processes are not stationary. Nonetheless, one can still argue that the quantity $f_n^\alpha(M)$ converges to $1/N_t$ by an argument that uses more of the structure of the PINS, PT, and INS schemes. Here we give just a sketch of the argument.

When put in the form of transition kernels on (S,Y), there is a stationary distribution μ, which satisfies detailed balance with respect to each of the different transition kernels associated with any of the PINS, PT, and INS schemes, and also this is the only such stationary distribution under all the kernels associated with a given scheme. (It is in fact the symmetrized version of the joint stationary distribution on the $N_t$ particles, symmetrized over all temperature coordinate associations (c.f. Eq. (3.6) of Ref. (21).) It follows from the explicit form of μ that the marginal distribution on S is uniform on $\mathcal{P}\{N_t\}$. Using relative entropy with respect to μ as a Lyapunov function[32], one can show that, so long as each kernel associated with a given scheme is used infinitely often, the distribution μ(m) of the process after m algorithmic steps converges to μ as m → ∞. Denote

$$g_s(M) = \frac{1}{M}\sum_{m=1}^{M}\mathbf{1}_{S(m),s}$$

(B.3)

where

$$\mathbf{1}_{S(m),s} = \begin{cases} 1, & S(m) = s \\ 0, & otherwise \end{cases}.$$

(B.4)

Using a standard martingale argument on can show that $g_s(M)$ and the average (over $m = 1,...,M$) of the measure placed on s by the marginal distribution of μ(m) on S are asymptotically the same, and therefore $g_s^\beta(M) \to 1/|\mathcal{P}\{N_t\}|$ as $M \to \infty$. The statement given in Section II for the equal occupancy of temperatures can then be obtained by summing over all permutations where a given coordinate is fixed.

***Appendix C:***  The present discussion summarizes the design and construction of the PINS computational ensemble for studies of Section III.  In general, the temperature range of the computational ensemble is dictated by the system and phenomenology under investigation.  The lowest computational temperature is taken to be the lowest of physical interest while the highest temperature is typically chosen to be large enough to assure a proper sampling.  As a practical matter, the choice of the highest temperature amounts to selecting a value such that the simulation model in question exhibits a liquid-like behavior or one for which even conventional Metropolis methods prove adequate.  Beyond the selection of the overall temperature range, the general design of the tempering ensemble is trade-off between the desire to minimize computational effort (smaller number of temperatures) and the desire to facilitate information flow (larger number of temperatures).  The low and high-temperature limits for the $Ar_{38}$ ensemble of Section III, 10 K and 30 K, respectively, are chosen to bracket the temperature of the major heat capacity peak for the $Ar_{38}$ system,[18] a feature that marks the boundary between solid-like and liquid-like behavior for the cluster.

In principle, once the total number, low, and high values are chosen, one can utilize the asymptotic decay rate of the occupation entropy to its known limit to optimize the selection of the remaining "interior" temperatures in the computational ensemble.  In practice, we have found for the applications considered to date that the performance of the PINS approach is sufficiently robust that such a precise temperature selection process is unnecessary.

The approach we have adopted for the selection of the interior ensemble temperatures in the present work is based on the desire to enhance the rate of information flow induced by the PINS symmetrization.  In practical terms, this translates into selecting the interior temperatures in such a way that the statistical importance of the various permutations that arise within the symmetrization blocks be as widely distributed as possible.  In the notation of Ref. (21), if one of the blocks in the dual-chain PINS approach contains N total temperatures, there are N! possible permutations of coordinate and temperature sets involved.  As part of the PINS sampling process, the statistical weights of these permutations (c.f. Eq. (3.5) of Ref. 21) relative to the total, $\{\rho_n\}$, n=1,N! are computed. For a given set of coordinates the entropies associated with these weights, defined for each of the symmetrized blocks as

$$S_\rho = -\sum_{n=1}^{N_t} \rho_n \ln[\rho_n]$$

(C.1)

thus provides a measure of the dispersal of the statistical weight across the set of permutations. For example, a value of zero for $S_\rho$ corresponds to all of the statistical weight being concentrated in a single permutation, whereas the maximum value of ln(N!) signifies that each permutation carries the same uniform statistical weight.

In the case of the 66-temperature $Ar_{38}$ ensemble of Section III we have chosen the number and distribution of temperatures in the computational ensemble so that the $S_\rho$ values for the various temperature blocks that are produced represent an appreciable and uniform fraction of the maximum values possible. The explicit temperatures for this ensemble are listed in Table II. Table III shows the average of the $S_\rho$ values produced by this ensemble during the two million move simulation discussed in Section III that was initiated from the global minimum configuration. Although not shown, the corresponding results for the simulation initiated using the lowest-lying icosahedral local minimum configuration are statistically equivalent those of Table III.

*Table I*

Observed fractions of total moves ($f_n$) spent at each of the ensemble temperatures by two, three-temperature $Ar_{13}$ parallel tempering (PT) and partial swapping (PINS) simulations. In both ensembles, $T_1 = 30$ K and $T_3 = 40$ K, while in one ensemble $T_2 = 35$ K and in the other $T_2 = 39$ K.

| $T_n$ | $f_n$ (PT) | $f_n$ (PINS) | $T_n$ | $f_n$ (PT) | $f_n$ (PINS) |
|---|---|---|---|---|---|
| 30 | 0.3314 | 0.3317 | 30 | 0.3342 | 0.3348 |
| 35 | 0.3330 | 0.3329 | 39 | 0.3334 | 0.3324 |
| 40 | 0.3356 | 0.3354 | 40 | 0.3324 | 0.3328 |

*Table II*

The temperatures used in the PINS $Ar_{38}$ computational ensemble.

| n | $T_n$ | n | $T_n$ | n | $T_n$ |
|---|---|---|---|---|---|
| 1 | 10.0000 | 23 | 15.7142 | 45 | 21.6364 |
| 2 | 10.2597 | 24 | 15.9740 | 46 | 22.0000 |
| 3 | 10.5195 | 25 | 16.2337 | 47 | 22.3637 |
| 4 | 10.7792 | 26 | 16.4935 | 48 | 22.7273 |
| 5 | 11.0390 | 27 | 16.7532 | 49 | 23.0909 |
| 6 | 11.2987 | 28 | 17.0130 | 50 | 23.4546 |
| 7 | 11.5585 | 29 | 17.2727 | 51 | 23.8182 |
| 8 | 11.8182 | 30 | 17.5324 | 52 | 24.1819 |
| 9 | 12.0779 | 31 | 17.7922 | 53 | 24.5455 |
| 10 | 12.3377 | 32 | 18.0519 | 54 | 24.9091 |
| 11 | 12.5974 | 33 | 18.3117 | 55 | 25.2727 |
| 12 | 12.8572 | 34 | 18.5714 | 56 | 25.6364 |
| 13 | 13.1169 | 35 | 18.8312 | 57 | 26.0000 |
| 14 | 13.3767 | 36 | 19.0909 | 58 | 26.3636 |
| 15 | 13.6364 | 37 | 19.3506 | 59 | 26.8182 |
| 16 | 13.8961 | 38 | 19.6104 | 60 | 27.2727 |
| 17 | 14.1559 | 39 | 19.8701 | 61 | 27.7272 |
| 18 | 14.4156 | 40 | 20.1299 | 62 | 28.1818 |
| 19 | 14.6753 | 41 | 20.3896 | 63 | 28.6364 |
| 20 | 14.9350 | 42 | 20.6494 | 64 | 29.0909 |
| 21 | 15.1948 | 43 | 20.9091 | 65 | 29.5454 |
| 22 | 15.4545 | 44 | 21.2727 | 66 | 30.0000 |

## Table III

Average of $S_\rho$ values, $\langle S_\rho \rangle$, for $Ar_{38}$ PINS simulation obtained using the computational ensemble shown in Table II. For reference, the maximum values for $S_\rho$ correspond to $\ln(3!) = 1.792$ and $\ln(6!) = 6.579$.

| Chain-1 | | Chain-2 | |
|---|---|---|---|
| T-range | $\langle S_\rho \rangle$ | T-range | $\langle S_\rho \rangle$ |
| 1 - 3 | 1.751 | 1 - 6 | 6.256 |
| 4 - 9 | 6.284 | 7 - 12 | 6.307 |
| 10 - 15 | 6.327 | 13 - 18 | 6.345 |
| 16 - 21 | 6.362 | 19 - 24 | 6.375 |
| 22 - 27 | 6.384 | 25 - 30 | 6.383 |
| 28 - 33 | 6.371 | 31 - 36 | 6.352 |
| 34 - 39 | 6.341 | 37 - 42 | 6.351 |
| 40 - 45 | 6.324 | 43 - 48 | 6.285 |
| 46 - 51 | 6.348 | 49 - 54 | 6.385 |
| 52 - 57 | 6.407 | 55 - 60 | 6.393 |
| 58 - 63 | 6.356 | 61 - 66 | 6.371 |
| 64 - 66 | 1.768 | | |

*Figures:*

Figure 1: Occupation traces for three-temperature PT (a,b) and PINS (c,d) simulations for $Ar_{13}$. $T_1 = 30$ K and $T_3 = 40$ K for all simulations, while $T_2 = 35$ K in simulations (a) and (c) and 39 K in (b) and (d).

Figure 2: Plots of the average number of moves required for a round-trip transit of the computational ensemble, $<n_{rt}>$ as function of $T_2$, for extended versions of the three-temperature $Ar_{13}$ simulations of the type in Fig. (1).

Figure 3: Approach of $S_f(n_{move})$(c.f. Eq. (2.3)) to its uniform limiting value for the three-temperature PINS and PT simulations of $Ar_{13}$ used in Table I and described in the text. $T_1 = 30$ K, $T_3 = 40$ K, $T_2 = 35$ K or 39 K.

Figure 4: Plot of $ln(S_{max} - S_f)$ for the three-temperature $Ar_{13}$ results of Fig. (3).

Figure 5: Plots of $C(s)$ (c.f. Eq. (2.4)) for the three-temperature $Ar_{13}$ simulations of Fig. (3).

Figure 6: A portion of occupation traces for 66-temperature $Ar_{38}$ PINS (black) and PT(red) simulations discussed in the text. The vertical axis denotes the temperature index (1-66) as a function of the number of moves in the simulation.

Figure 7: A histogram of the PINS occupation trace shown in Fig. 6 showing the number of times the various temperature indices are visited, $M(n)$, as a function of n.

Figure 8: A plot of $S_f(n_{move})$ obtained for $Ar_{13}$ using PINS and PT methods for the various 24-temperature ensembles described in the text. The apparent "break" in the PINS-24a results occurs at an $S_f$ value of roughly 2.5, a value that corresponds to an active number of temperatures ($N_a$) of approximately 12.

Figure 9: Plot of $ln(S_{max} - S_f)$ for results in Fig. (8).

Figure 10: $C(s)$ for the PINS $Ar_{13}$ results obtained using the three, 24-temperature ensembles described in the text. PT results for ensemble-c are shown for comparison (dashed line near top of plot).

Figure 11: Brief portions of occupation traces for PINS (top panel) and PT (bottom panel) simulations for $Ar_{13}$ obtained using 24-temperature ensemble-c (see text for details).

Figure 12: Plots of the occupation entropy, $S_f(nmove)$, for 66-temperature PINS and PT simulations of the $Ar_{38}$ system. The two PINS results correspond to simulations that are initiated in the global minimum geometry (black curve) or lowest-lying icosahedral minimum (red curve). The limiting $S_f$ value of $\ln(66)$ is shown for reference.

Figure 13: Plot of $\ln(S_{max} - S_f)$ for results in Fig. (12).

Figure 14: A plot of $<Q_4(T)>$ for the $Ar_{38}$ cluster obtained by the PINS simulations described in the text. Results in black (red) are obtained using a simulation initialized using the fcc global minimum (icosahedral local minimum) structure. For clarity and as an aid in comparing the two simulations $<Q_4(T)>$ only values for every other (every fourth) temperature are shown for the fcc (icosahedral) results.

Figure 15: $Q_4$ values for an extended portion of the global minimum inititated occupation trace of Fig. (6).

Figure 16: A history of the number of configurations (out of 66) in the two PINS $Ar_{38}$ simulations described in the text for which $Q_4 \leq 0.09$.

Figure 17: Block averages of $Q_4$ for $Ar_{38}$ for $T_1 = 10$ K, $T_2 = 14.9350$ K for the PINS simulations described in the text.

Figure 18: Shown are the $Q_4$ values for $T = 22.727$ K for a short, post warm up portion of the icosahedral minimum initiated $Ar_{38}$ PINS simulation. Compare with Fig. (11) of Ref. (29).

Figure 1:

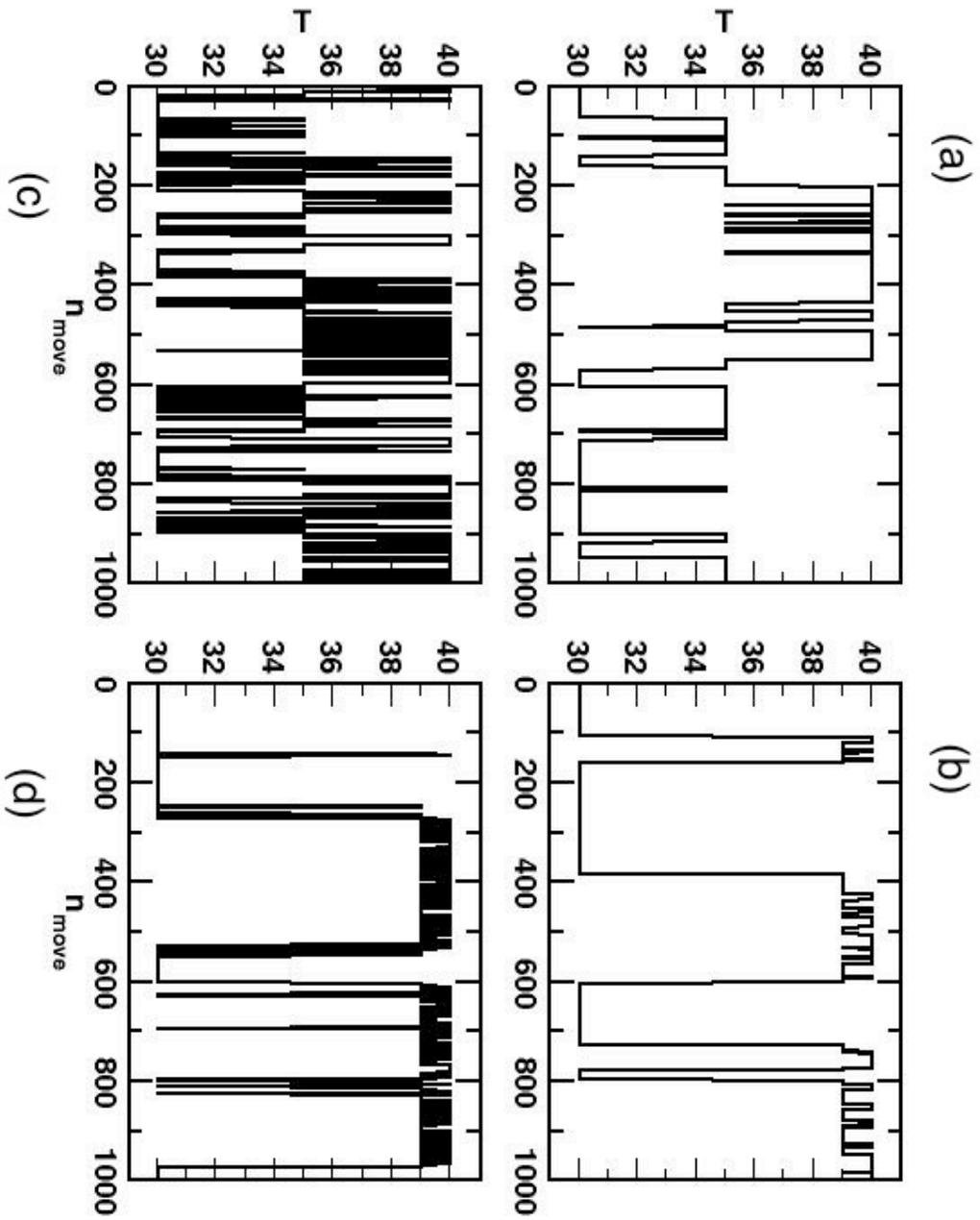

Figure 2:

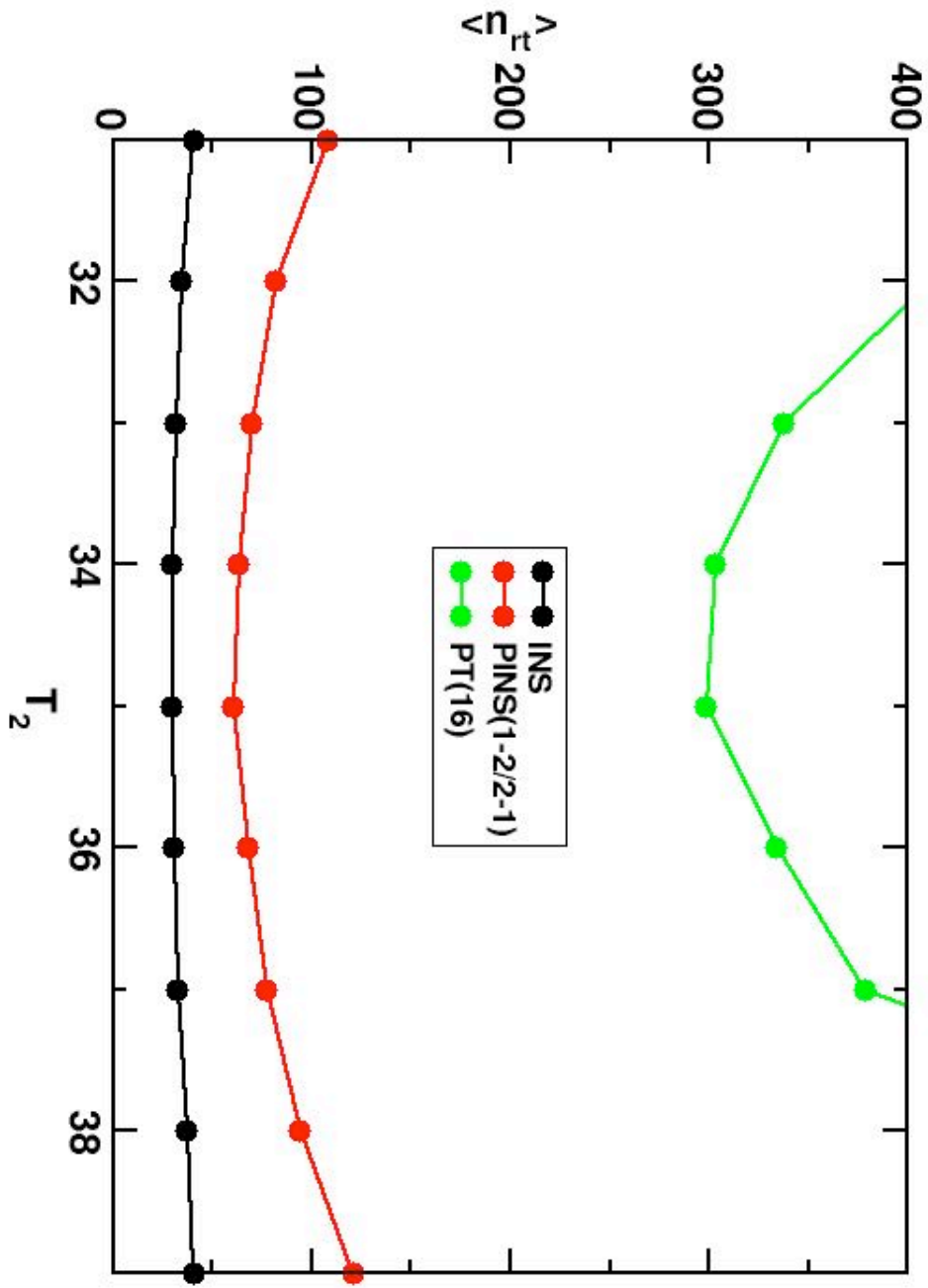

Figure 3:

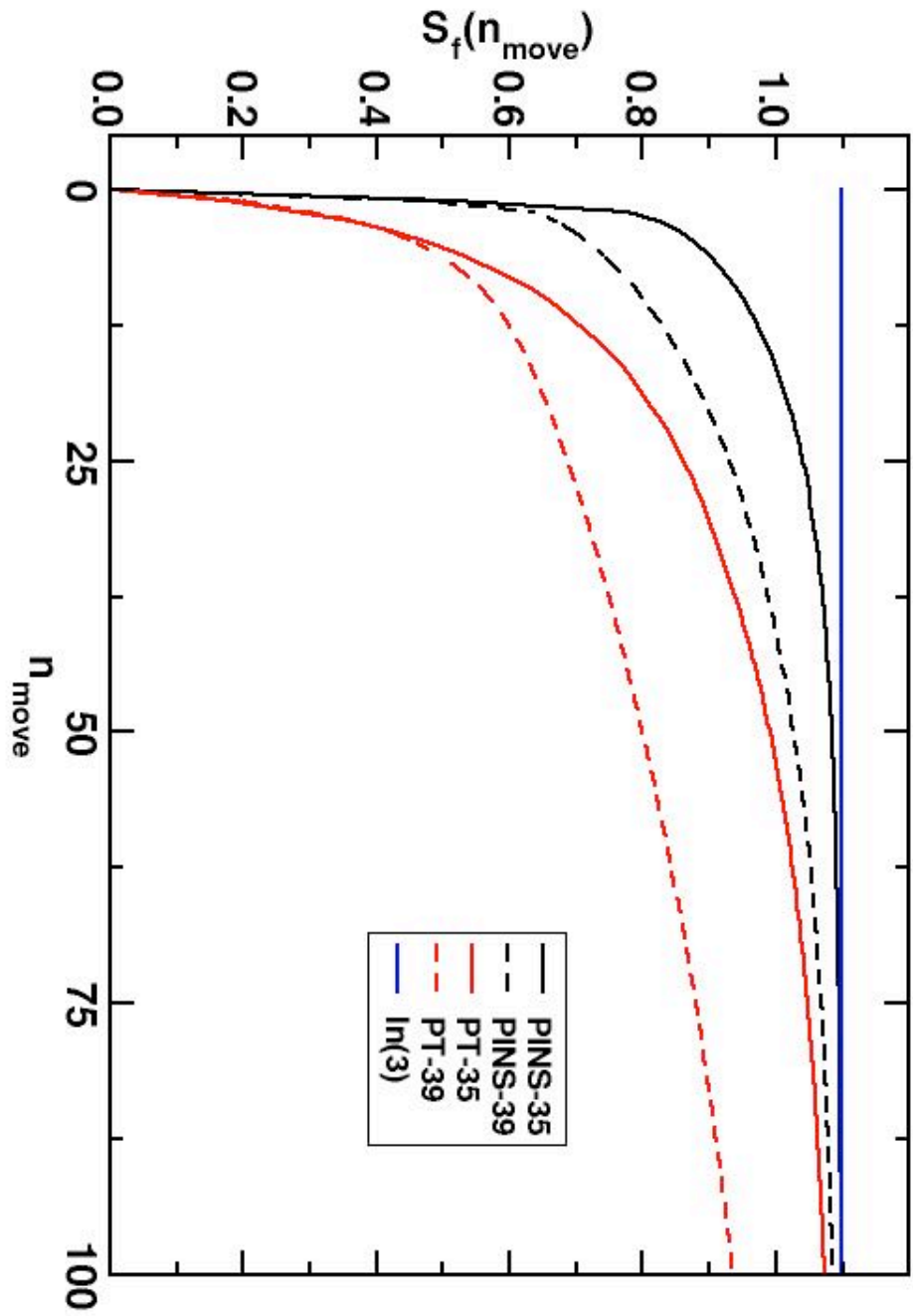

Figure 4:

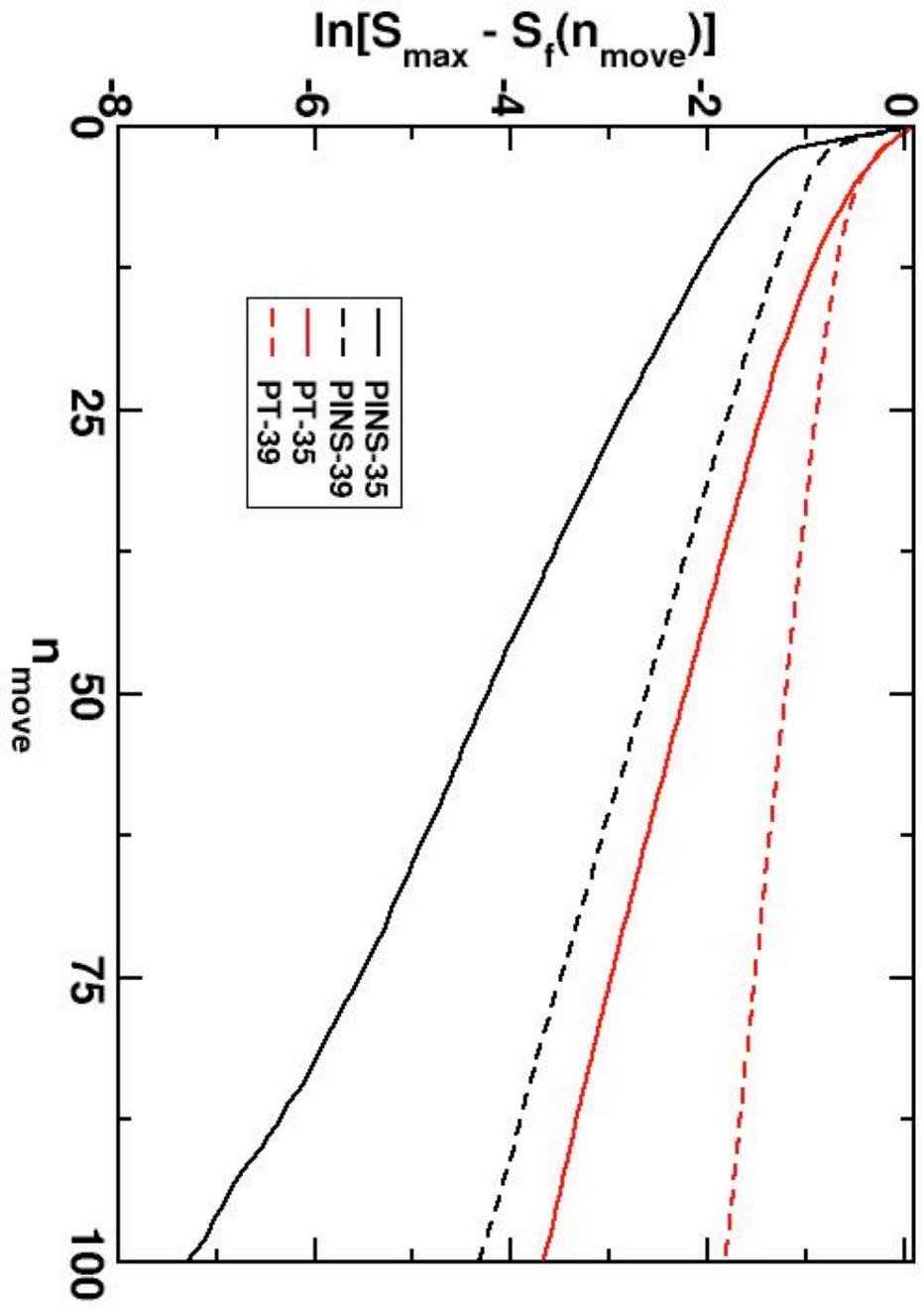

Figure 5:

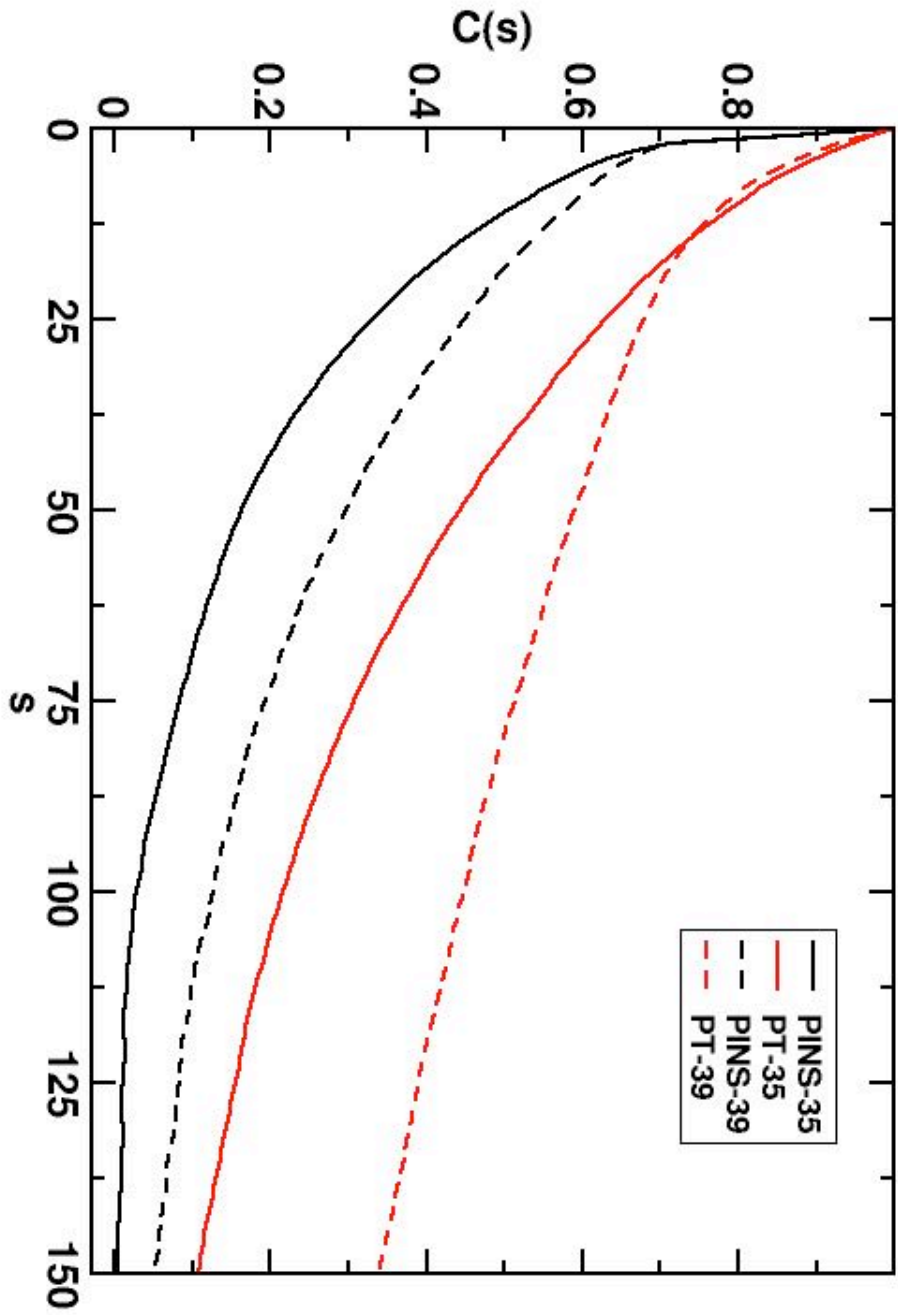

Figure 6:

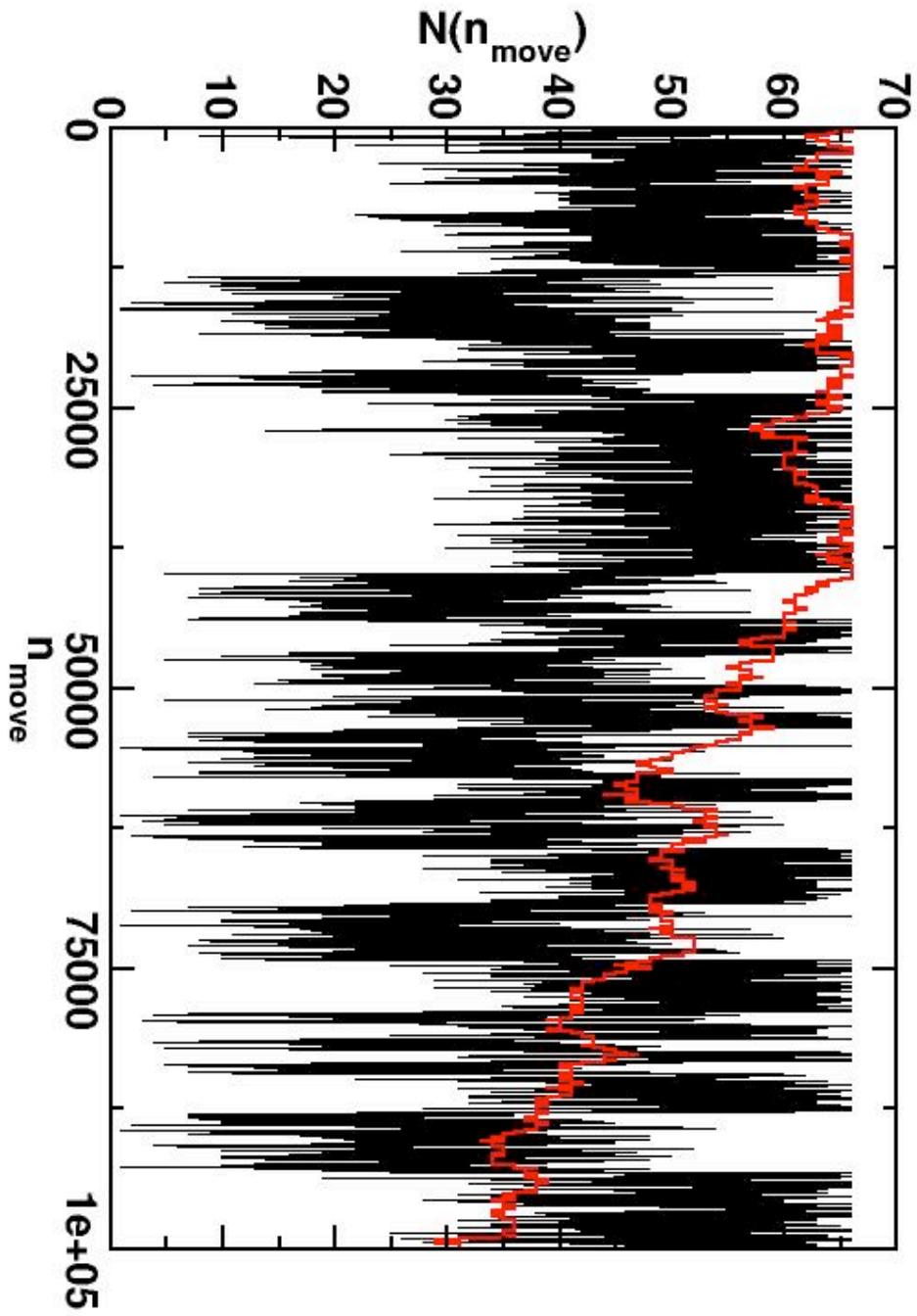

Figure 7:

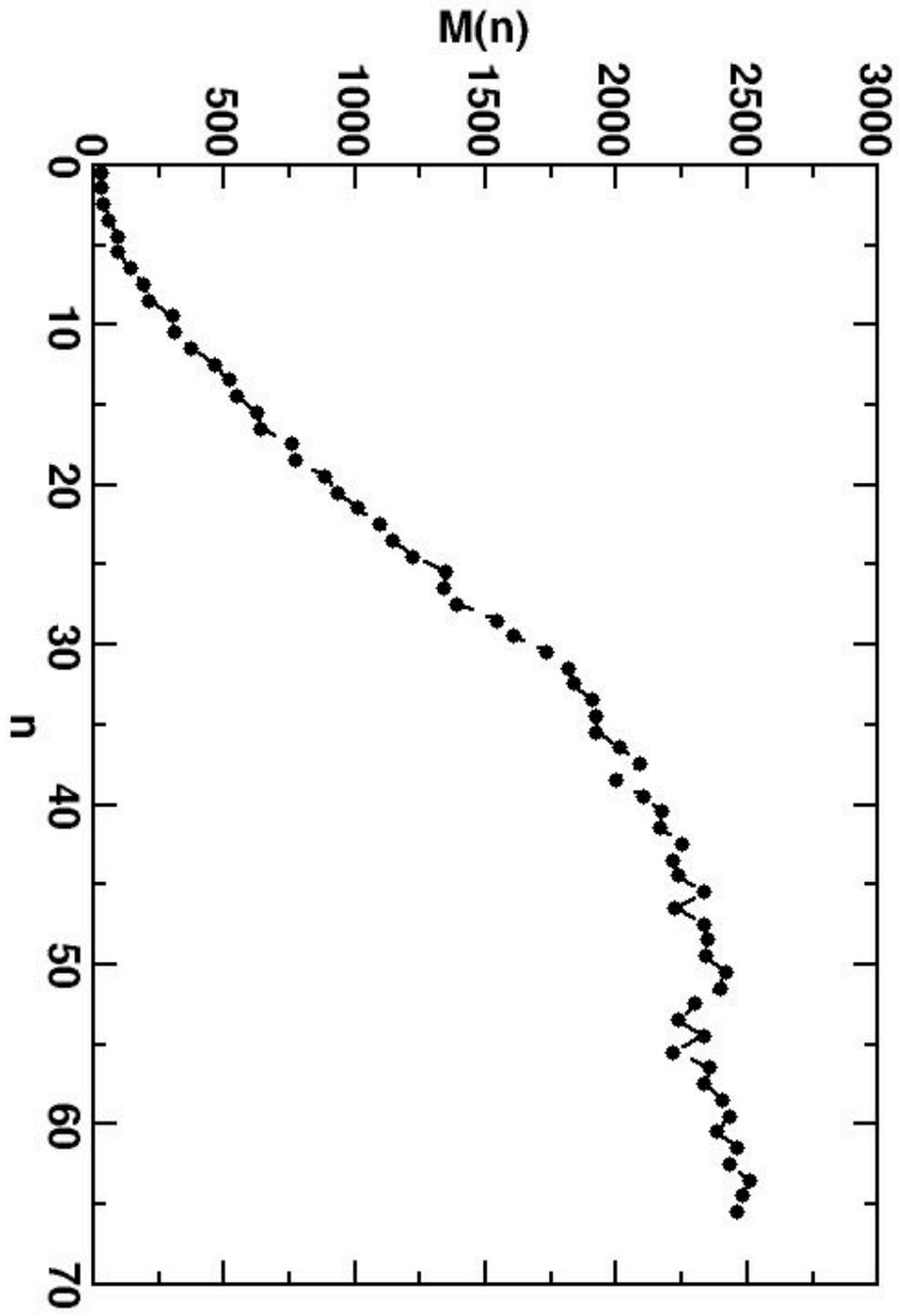

Figure 8:

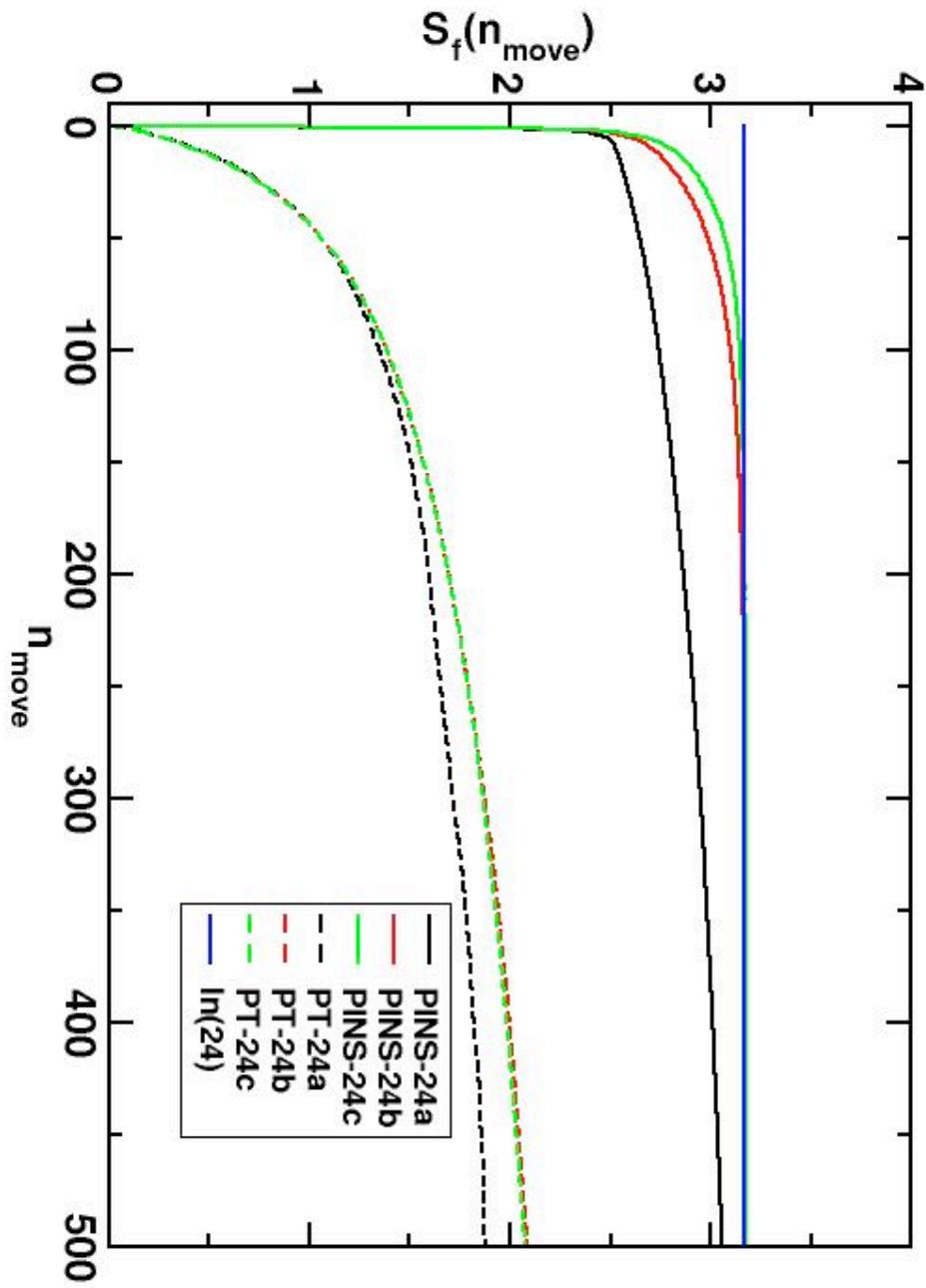

Figure 9:

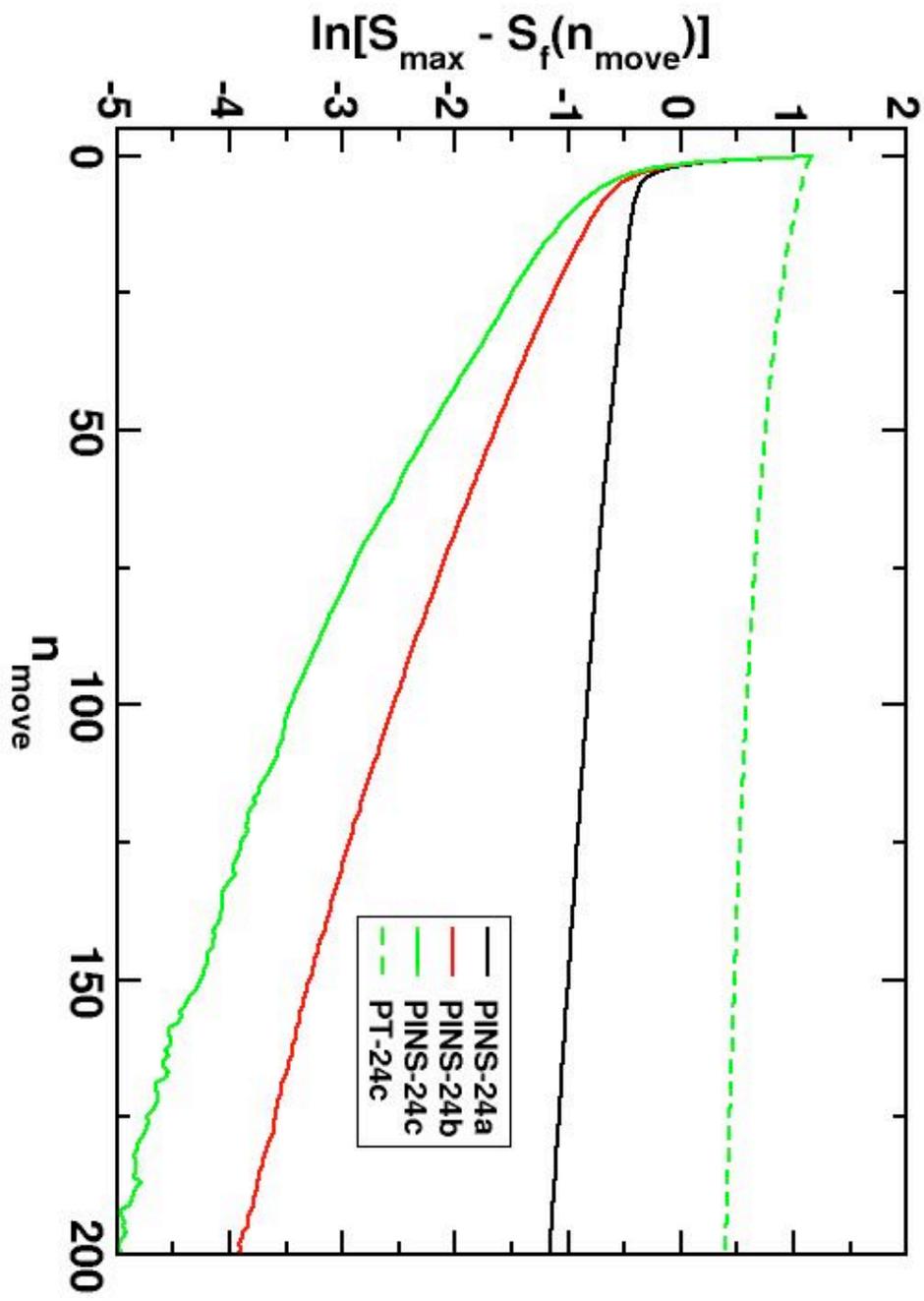

Figure 10:

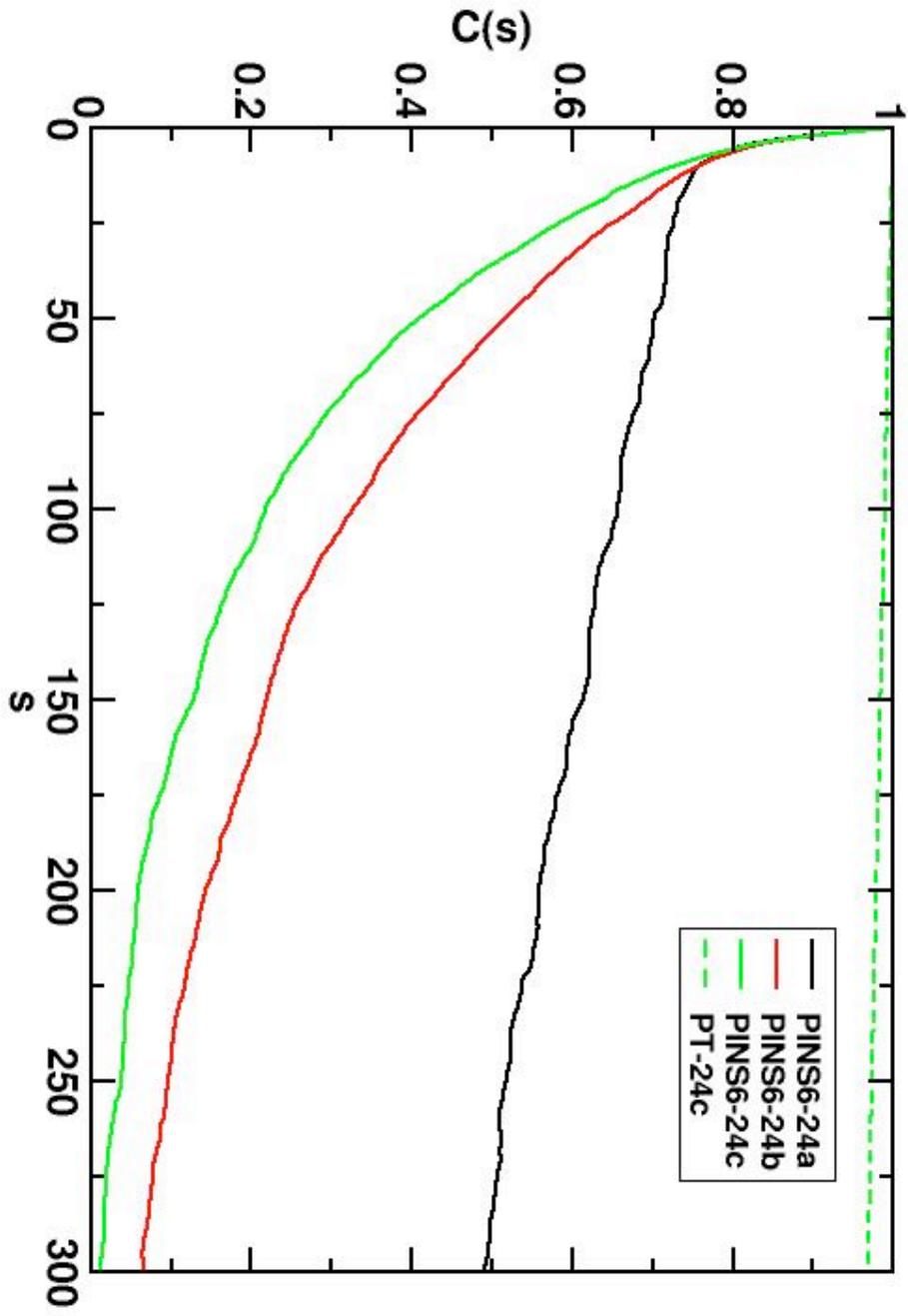

Figure 11:

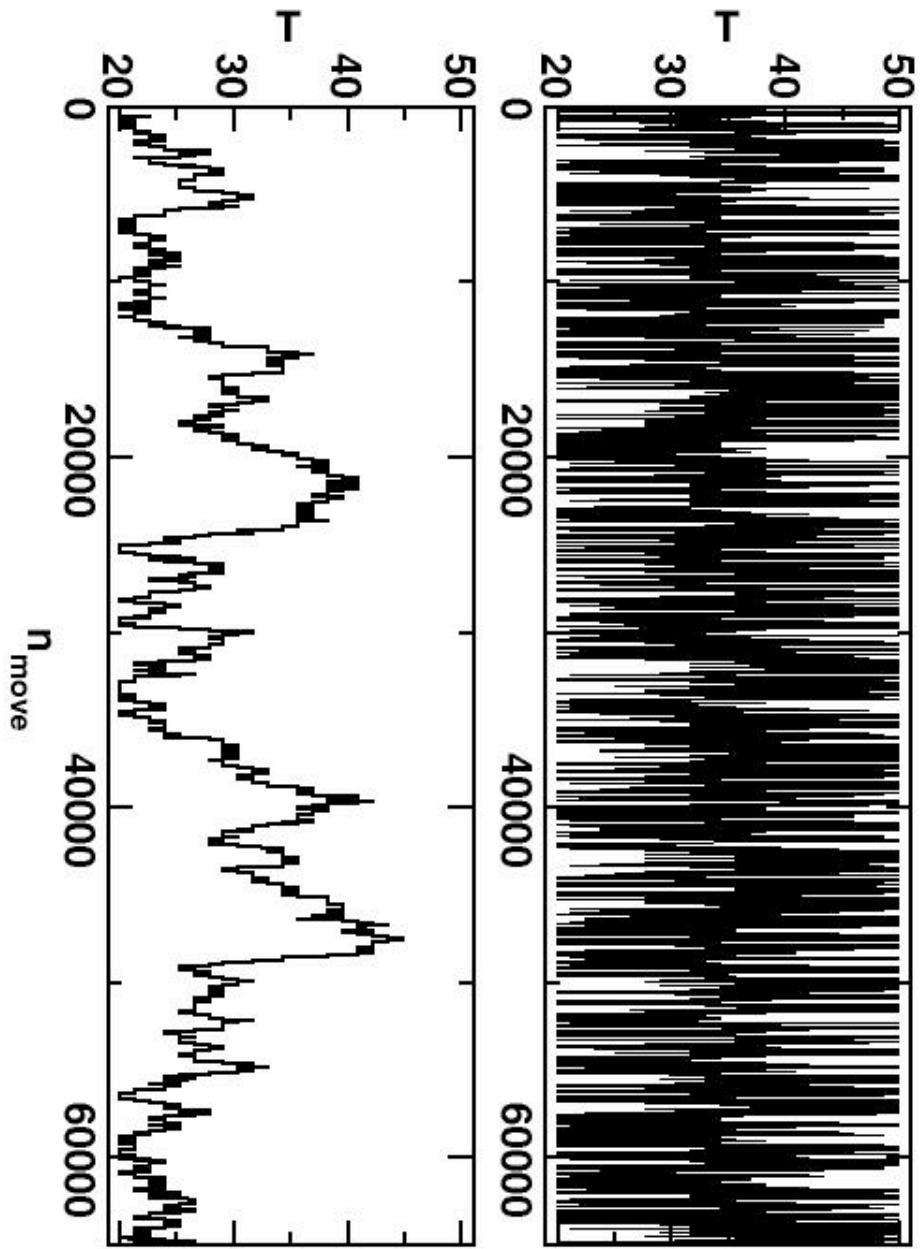

Figure 12:

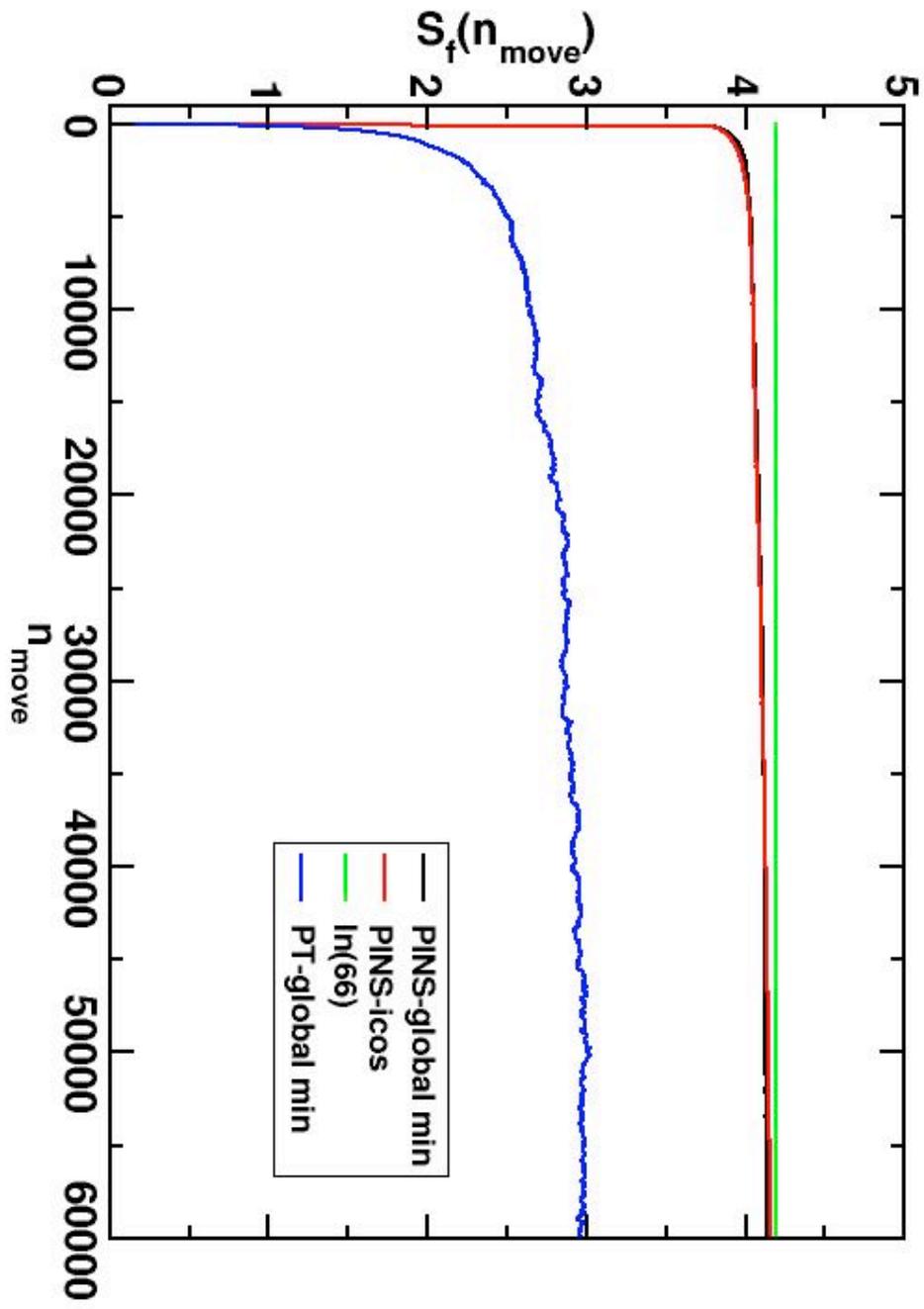

Figure 13:

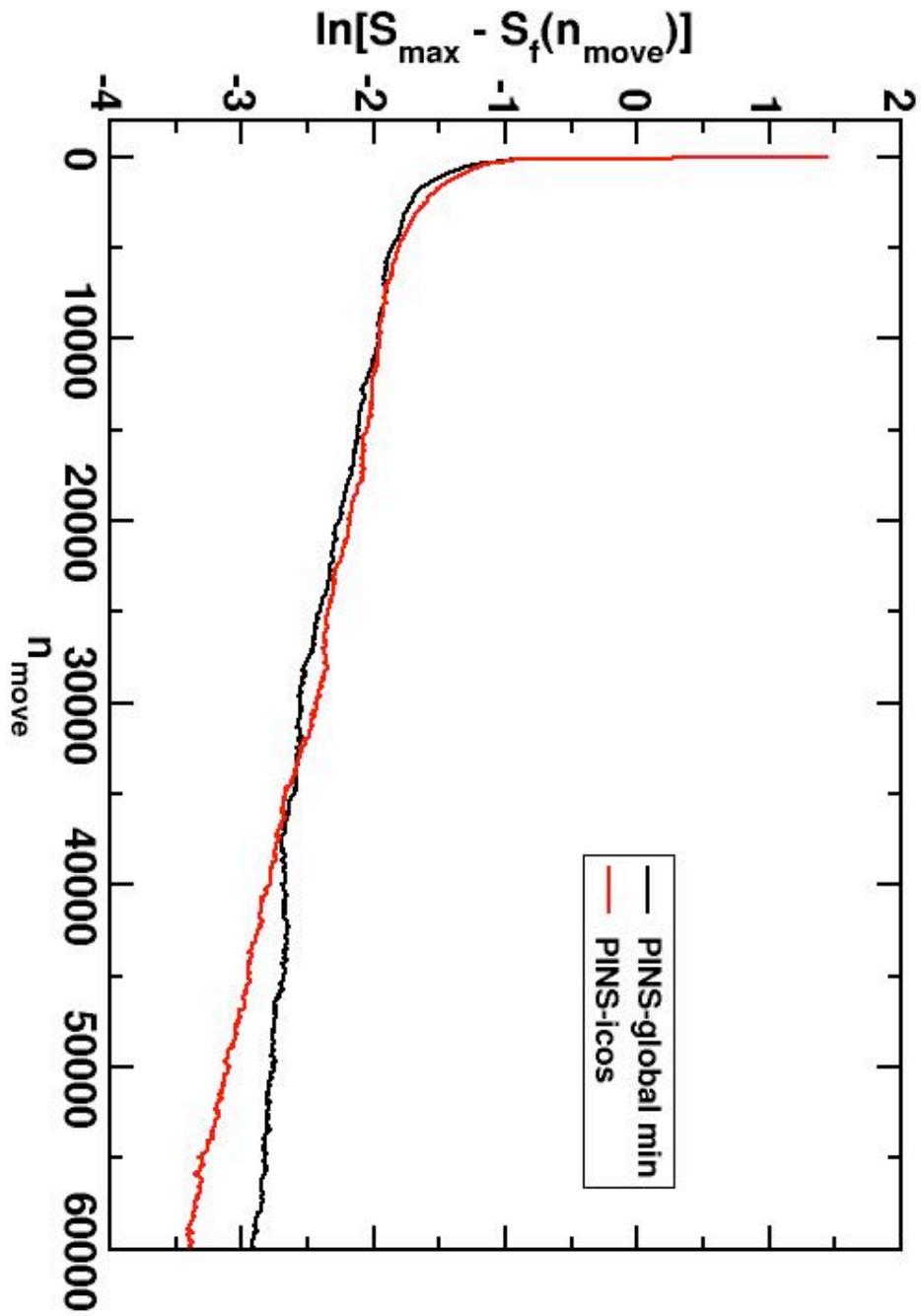

Figure 14:

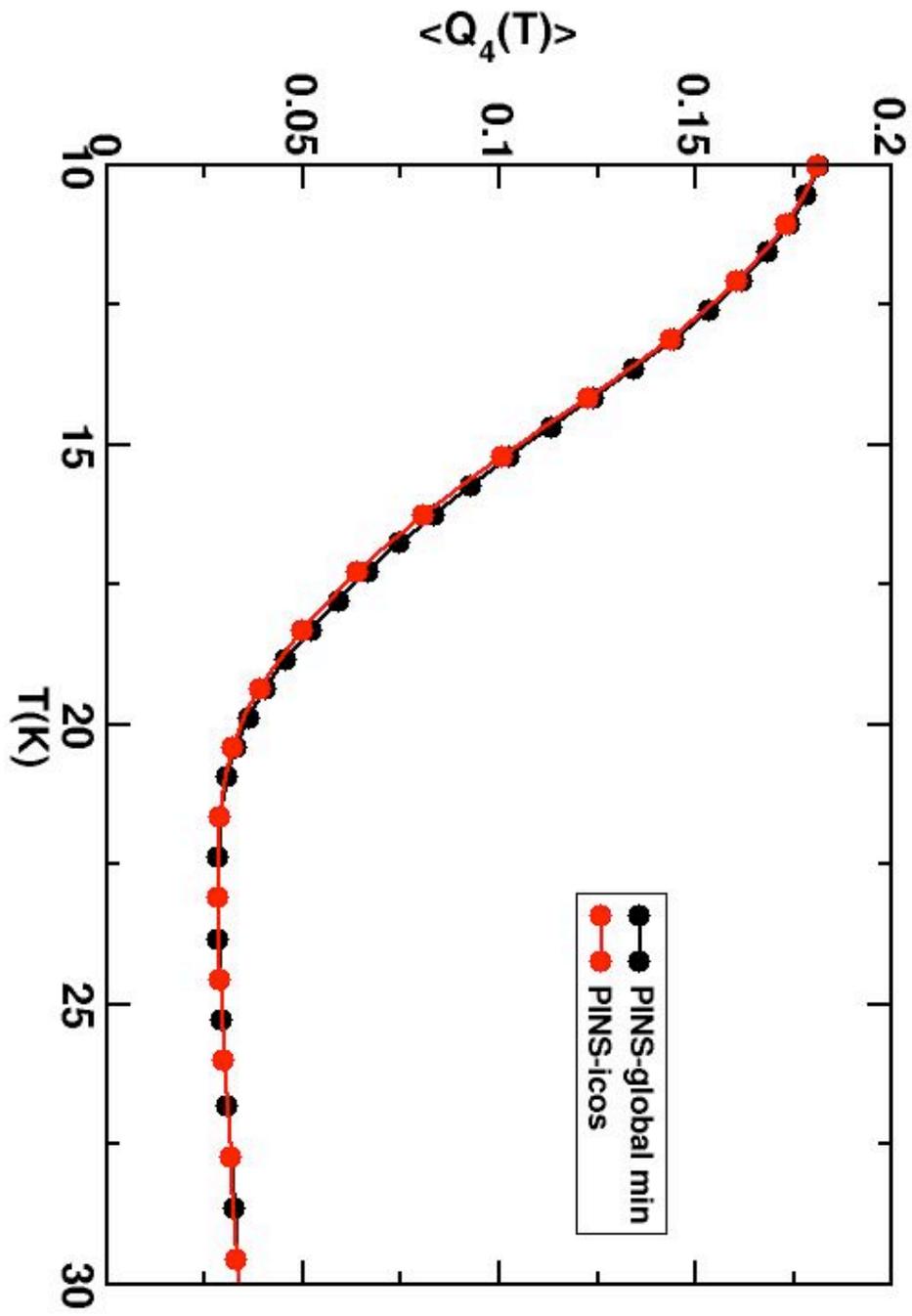

Figure 15:

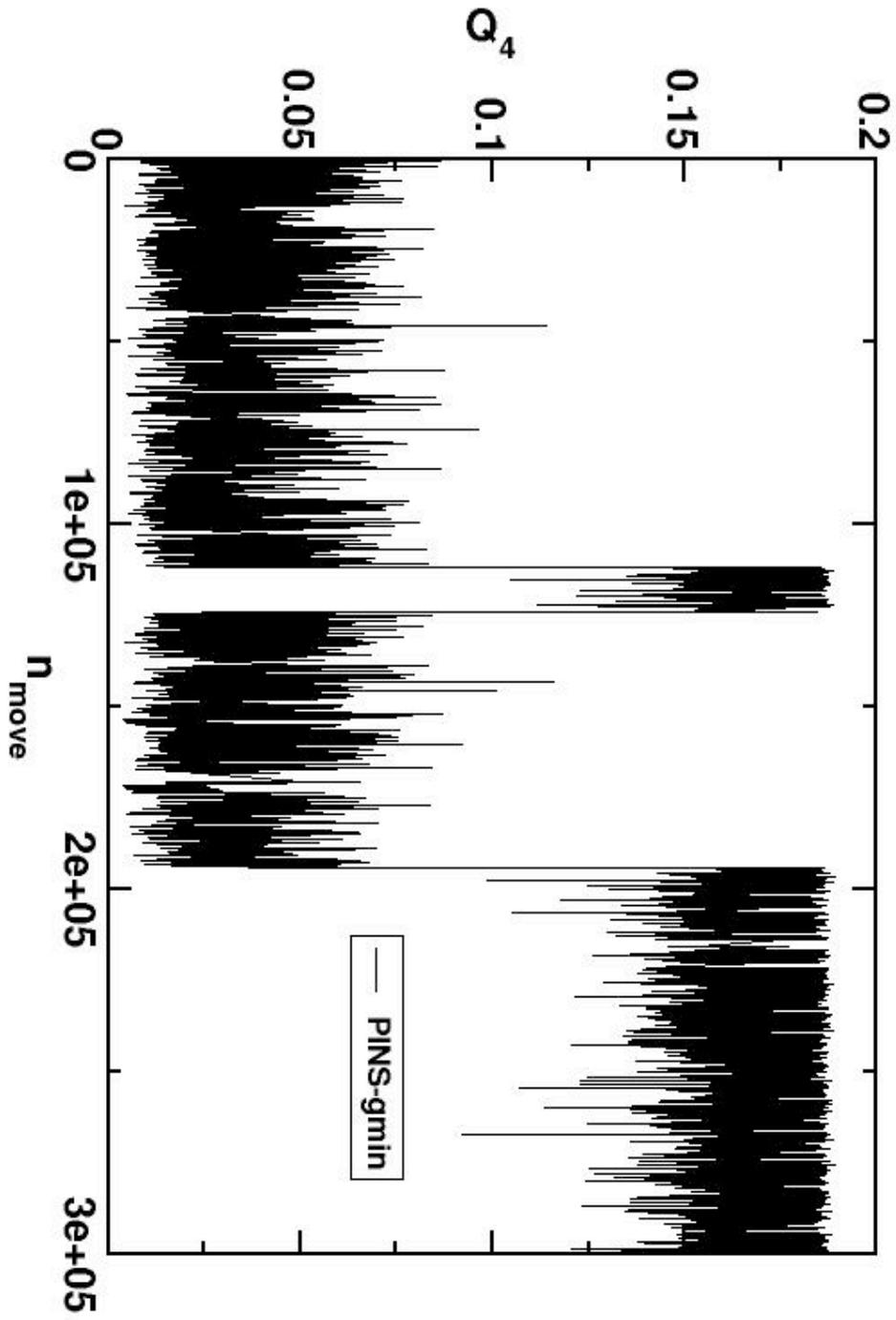

Figure 16:

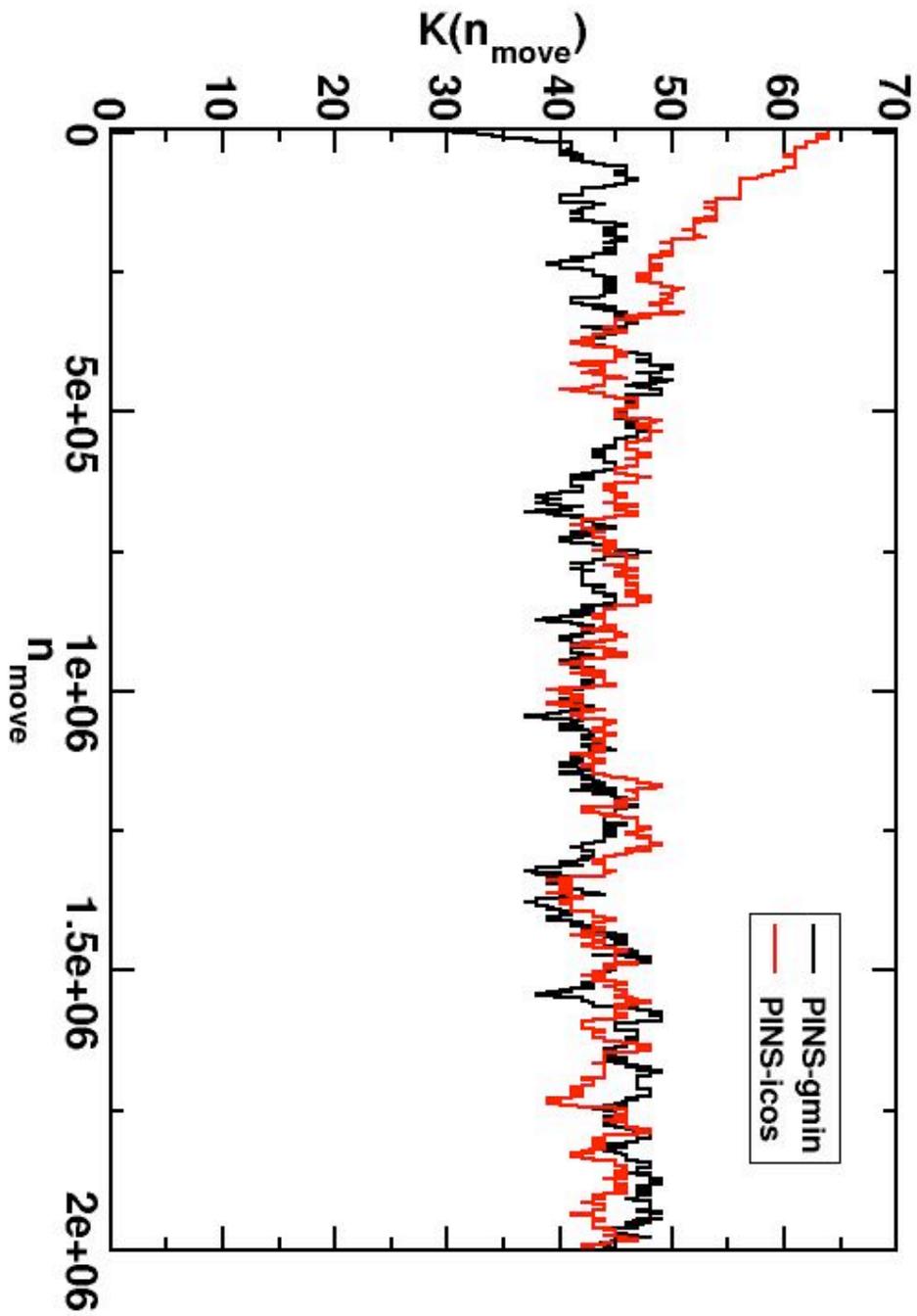

Figure 17:

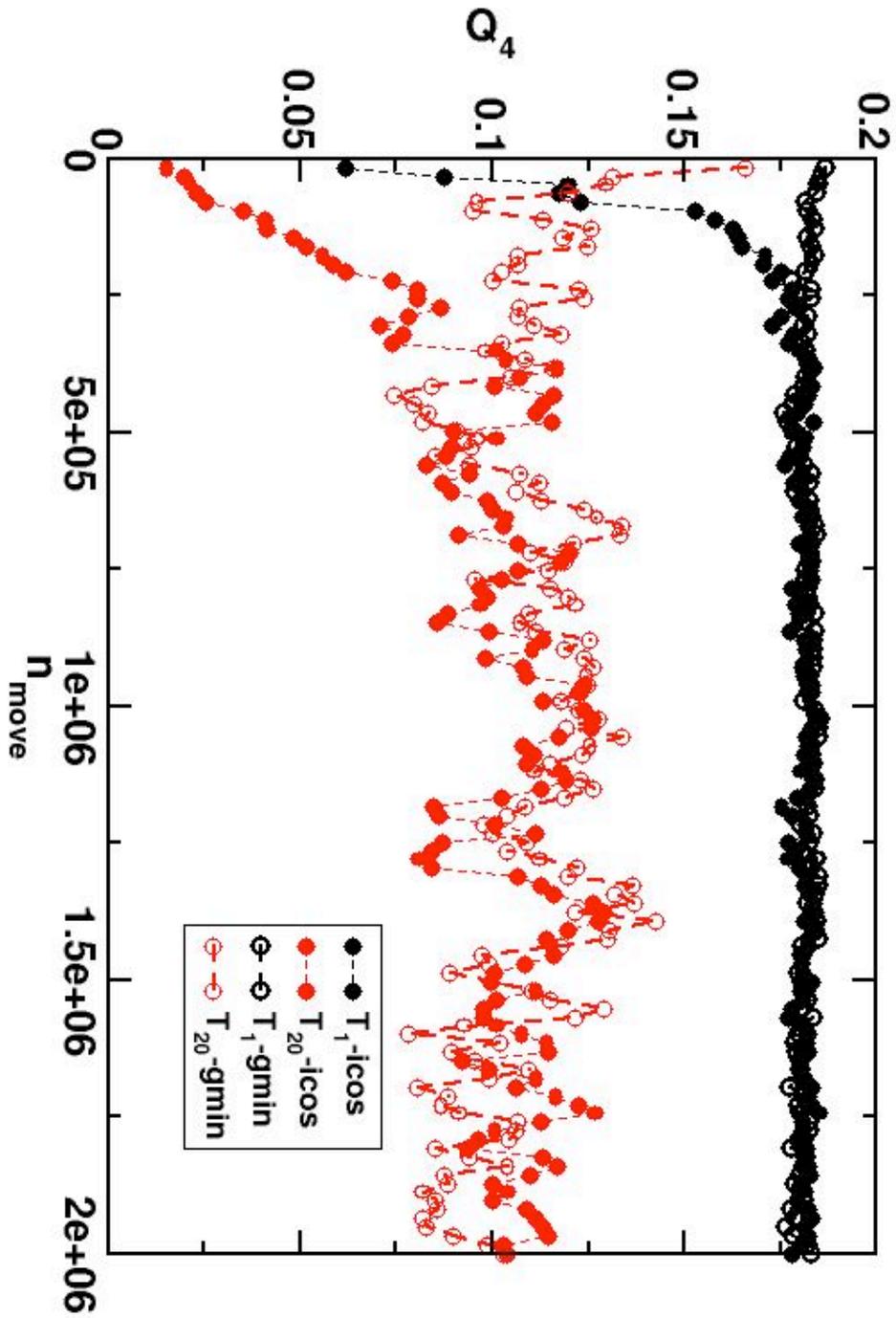

Figure 18:

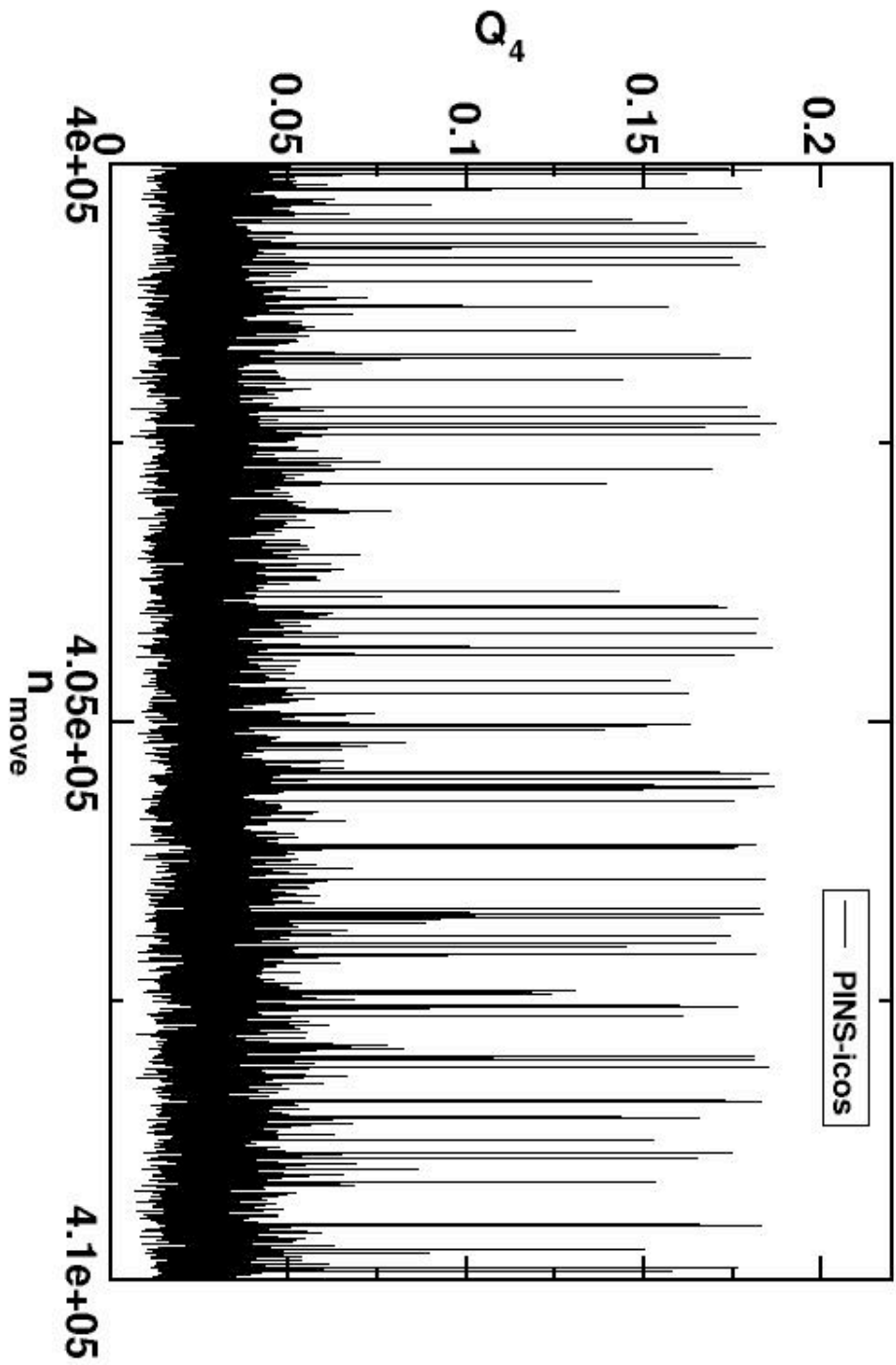